\documentclass[12pt]{article}
\usepackage{amsthm,amsmath,latexsym,amssymb,amsfonts,amscd}
\usepackage{graphics,lscape,fancyhdr,array,stmaryrd,euscript,wrapfig}
\usepackage{empheq}
\usepackage{verbatim,slashed}
\numberwithin{equation}{section}
\usepackage{hyperref,setspace}
\usepackage[numbers,sort&compress]{natbib}
\usepackage[nottoc]{tocbibind}
\usepackage{tikz-cd}
\usepackage[all]{xy}
\usepackage{tikz}
\usetikzlibrary{arrows.meta}
\usetikzlibrary{decorations.pathmorphing}
\usepackage{multirow}

\usepackage{rotating}

\numberwithin{equation}{section}
\newcommand{\pl}{\partial}
\newcommand{\plb}{\bar{\partial}}
\newcommand{\fud}[2]{{}^{#1}{}_{#2}\,}
\newcommand{\fdu}[2]{{}_{#1}{}^{#2}\,}

\newcommand{\ga}{A}
\newcommand{\gb}{B}

\newcommand{\gad}{{A'}}
\newcommand{\gbd}{{B'}}

\newcommand{\bry}{{{\bar{y}}}}

\newcommand{\brk}{{{\bar{k}}}}

\newcommand{\action}[2]{{\left\langle\vphantom{#2}#1\,\right|\left.\vphantom{#1}#2\right\rangle}}
\newcommand{\scalar}[2]{{\left\langle\vphantom{#2}#1\right.;\left.\vphantom{#1}#2\right\rangle}}

\newcommand{\besubeqs}{\begin{subequations}}
\newcommand{\esubeqs}{\end{subequations}}

\newcommand{\sign}{\operatorname{sign}}

\begin{document}
\hfill
\vskip 0.01\textheight
\begin{center}
{\large\bfseries 
Self-dual holography:\\
\vspace{0.2cm}
four-point AdS/CFT correlators in higher-spin gravity}

\vspace{0.4cm}

\vskip 0.03\textheight
\renewcommand{\thefootnote}{\fnsymbol{footnote}}
Evgeny \textsc{Skvortsov}\footnote{Also affiliated with Lebedev Institute of Physics.}  \& Richard van Dongen

\vskip 0.03\textheight

{\em Service de Physique de l'Univers, Champs et Gravitation, \\ Universit\'e de Mons, 20 place du Parc, 7000 Mons, 
Belgium}\\
\vspace*{5pt}

\renewcommand{\thefootnote}{\arabic{footnote}}
\end{center}

\vskip 0.02\textheight

\begin{abstract}
Self-dual theories, being UV-finite, should have their own AdS/CFT dualities. Higher-spin extensions of self-dual theories are attractive to simplify the CFT duals. The maximal self-dual theory is Chiral higher-spin gravity, which should be dual to a subsector of Chern--Simons matter theories. As a step toward establishing self-dual holography, we develop the Fefferman--Graham expansion and holographic dictionary for arbitrary spin self-dual theories. As an application, we derive bulk-to-bulk propagators and compute three- and four-point AdS/CFT correlators in a contraction of Chiral higher-spin gravity, which is a higher-spin extension of self-dual Yang--Mills theory.
\end{abstract}

\newpage
\tableofcontents
\newpage

\section{Introduction}
\label{sec:intro}
The main goal of this paper is to further develop self-dual holography \cite{Skvortsov:2018uru,Sharapov:2022awp,Jain:2024bza,Aharony:2024nqs,Chowdhury:2024dcy,Sharma:2025ntb,Skvortsov:2026gtq}, i.e. to identify the role of self-dual theories in (A)dS/CFT and celestial holography. Concretely, we concentrate on self-dual theories with massless fields of arbitrary spin in AdS${}_4$. The maximal such theory --- the ``M-theory'' of self-dual ones --- is chiral higher-spin gravity \cite{Metsaev:1991mt,Metsaev:1991nb, Ponomarev:2016lrm, Sharapov:2022awp, Sharapov:2022wpz}. Its spectrum has fields of all spins and it features all interactions compatible with self-duality.  

Self-dual theories are of great interest for many reasons, to mention a few: (a) they are consistent truncations of some bigger nonchiral theories and all solutions of self-dual theories are also solutions of the parent ones; (b) self-dual theories are integrable; (c) all amplitudes of self-dual theories are also amplitudes of their completions, i.e. self-dual theories allow one to compute some, but not all, amplitudes of more physical theories; (d) they do not have UV divergences; (e) they admit simple twistor formulations; (f) the parent theories can be represented as expansions over the self-dual sectors, which is richer than the usual perturbative expansion over a trivial vacuum. Therefore, self-dual theories provide a very useful approximation of the parent theories.  

\begin{figure}
    \centering
    \includegraphics[width=0.75\linewidth]{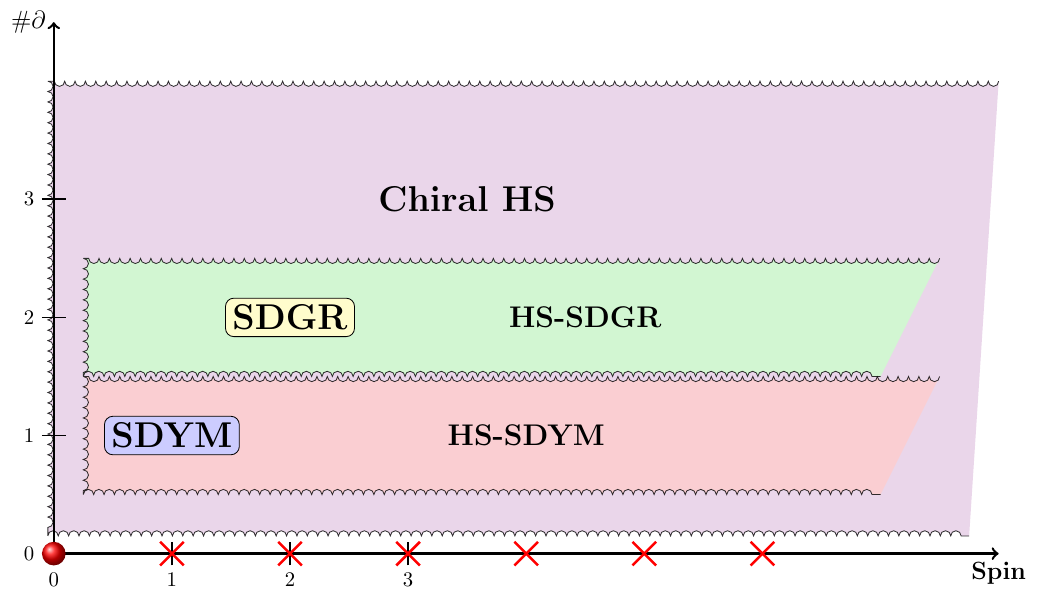}
    \caption{The ``map'' of self-dual theories. Along $x$ we plot spin and along $y$ we display the ``number of derivatives'' in cubic vertices. The smallest theories are SDYM and SDGR, which admit higher-spin extensions: HS-SDYM and HS-SDGR. The maximal self-dual theory is chiral higher-spin gravity. There are many intermediate cases \cite{Ponomarev:2017nrr,Serrani:2025owx} and the complete classification is not yet available. The $x$-line itself is forbidden as there are no vertices with vanishing total helicity except for the scalar self-interaction. }
    \label{fig:enter-label}
\end{figure}

Self-dual theories admit higher-spin extensions of various types, see the figure \ref{fig:enter-label} and \cite{Ponomarev:2016lrm, Ponomarev:2017nrr, Krasnov:2021nsq,Monteiro:2022xwq,Ponomarev:2024jyg,Serrani:2025owx} for more details that the figure can display. Provided we restrict ourselves to either gauge or gravitational interactions, there exist simple higher-spin extensions of self-dual Yang-Mills theory (SDYM) and self-dual gravity (SDGR) --- HS-SDYM and HS-SDGR. All self-dual theories, in particular those with genuine higher-spin higher-derivative nonabelian interactions, are contained in chiral higher-spin gravity, which thereby incorporates all interactions that are compatible with self-duality. The amplitudes of SDYM and SDGR are contained in the amplitudes of Chiral HiSGRA. Conversely, HS-SDYM and HS-SDGR compute some of the amplitudes of Chiral theory, which is the property we will use in the paper. 
Self-dual theories admit gauge- and Lorentz-invariant formulations, e.g. Chalmers--Siegel \cite{Chalmers:1996rq} and Krasnov \cite{Krasnov:2016emc} formulations for SDYM and SDGR, respectively. Covariant formulations for HS-SDYM and HS-SDGR were constructed in \cite{Krasnov:2021nsq}. However, the structure of the interactions is perhaps best exhibited in the light-cone gauge and in flat space. As is well-known, given any three helicities $\lambda_{1,2,3}$ such that $|\sum_i \lambda_i|>0$  (no discrimination of higher-spins), there is a unique cubic vertex $V_{\lambda_1,\lambda_2,\lambda_3}$ \cite{Bengtsson:1983pd,Bengtsson:1986kh} and the corresponding three-point amplitude reads \cite{Benincasa:2011pg}
\begin{align}\label{genericV}
   V_{\lambda_1,\lambda_2,\lambda_3}\Big|_{\text{on-shell}} \sim \begin{cases}
        [12]^{\lambda_1+\lambda_2-\lambda_3}[23]^{\lambda_2+\lambda_3-\lambda_1}[13]^{\lambda_1+\lambda_3-\lambda_2}\,, & \sum\lambda_i>0\,,\\
        \langle 12\rangle^{-\lambda_1-\lambda_2+\lambda_3}\langle23\rangle^{-\lambda_2-\lambda_3+\lambda_1}\langle13\rangle^{-\lambda_1-\lambda_3+\lambda_2}\,, & \sum\lambda_i<0\,.
   \end{cases}
\end{align}
For example, SDYM has a single interacting vertex $V_{+1,+1,-1}$, which is ``half'' of the Yang-Mills cubic vertex $V_{+1,+1,-1}+V_{-1,-1,+1}$. SDGR has only $V_{+2,+2,-2}$, which is again ``half'' of the cubic vertex of the Einstein-Hilbert action. HS-SDYM has all spins and $\sum_i \lambda_i=1$,\footnote{Let us fix the preferred value of $\Lambda=\sum \lambda_i$ to be positive. For every statement below there is a similar one with $\sum \lambda_i<0$. The number of (transverse) derivatives in a vertex in the light-cone gauge is $|\Lambda|$.} i.e. all vertices are single-derivative of the Yang-Mills type. HS-SDGR also has all spins but $\sum_i \lambda_i=2$, i.e. all interactions are two-derivative gravitational-type interactions. 

Chiral HiSGRA has all spins and all possible interactions with $\sum_i \lambda_i>0$: under quite general assumptions having a genuine higher-spin interaction forces all helicities to be present and also fixes the coupling constants to be \cite{Metsaev:1991mt,Metsaev:1991nb, Ponomarev:2016lrm}
\begin{align}\label{eq:magicalcoupling}
    V_{\text{Chiral}}&= \sum_{\lambda_1,\lambda_2,\lambda_3}  C_{\lambda_1,\lambda_2,\lambda_3}V_{\lambda_1,\lambda_2,\lambda_3}\,, && C_{\lambda_1,\lambda_2,\lambda_3}=\frac{\kappa\,(l_p)^{\lambda_1+\lambda_2+\lambda_3-1}}{\Gamma(\lambda_1+\lambda_2+\lambda_3)}\,.
\end{align}
As a result, there is a simple action for Chiral theory in the light-cone gauge in flat space. It is also possible to covariantize the theory at least at the level of equations of motion \cite{Sharapov:2022awp, Sharapov:2022wpz}. The equations of motion are smooth in the cosmological constant and Chiral theory exists also in $\text{(A)dS}_4$, which makes it an attractive model for holography since it is a perturbatively local and (supposed to be \cite{Skvortsov:2020wtf,Skvortsov:2020gpn}) UV-finite higher-spin gravity. 

Self-dual theories, being UV-finite, provide a natural playground for AdS/CFT correspondence (or for any other type of holography). Indeed, one needs a consistent theory of quantum gravity to have a complete model on the bulk side of the duality. (Super)gravity fails to be renormalizable and calls for a stringy completion, which complicates the model. Alternatively, self-dual theories are interesting toy models of quantum gravity. There is a good chance they can lead to nontrivial exactly soluble models of holography. There does not seem to be any other easily tractable framework in between self-dual theories and string theory that has propagating degrees of freedom, most notably the graviton. 

The last point requires more explanation since the statement has varied with time. When looking for simple models of the AdS/CFT correspondence it was initially noted that vector models (both the free and the critical one) should have ``simpler'' duals \cite{Klebanov:2002ja,Sezgin:2003pt,Leigh:2003gk}: conserved (higher-spin) currents present in the free vector model (or large-$N$ critical vector model) are dual to massless (higher-spin) fields and, hence, the AdS theory has to be a higher-spin gravity. Vector models are, in some sense, the smallest CFTs and the bulk dual has a finite number of fields for a given spin (say, one). Later, this idea was extended to Chern--Simons matter theories \cite{Giombi:2011kc}, i.e. to vector models with bosonic and/or fermionic matter coupled to Chern--Simons theory, and played an important role in establishing the three-dimensional bosonization duality \cite{Giombi:2011kc,Maldacena:2012sf,Aharony:2012nh}.

However, not every CFT has a quasi-classical and (perturbatively) local dual. The bulk coupling is of order $1/N$ for vector models and, hence, the dual admits a perturbative expansion. What prevents the dual from existing is the fact \cite{Bekaert:2015tva,Maldacena:2015iua,Sleight:2017pcz,Ponomarev:2017qab} that it has to be nonlocal beyond what field theory can tolerate.\footnote{\cite{Bekaert:2015tva} implies that the quartic vertex is as nonlocal as the exchange diagram, see, however, \cite{Neiman:2023orj}. } Therefore, at present it is impossible to construct the bulk dual of Chern--Simons matter theories and, hence, to compute correlation functions in the latter.\footnote{Some features of the bulk dual are captured by Vasiliev's equations \cite{Vasiliev:1990cm} and by the collective dipole approach \cite{Das:2003vw,deMelloKoch:2018ivk,Aharony:2020omh}. The former captures the symmetry aspects that are stable under nonlocal field-redefinitions, although deriving systematic predictions for interactions has remained challenging despite the efforts. Mathematically, it gives a certain $L_\infty$-algebra, rather than a field theory. The discrepancies show up already at the three-point level, see \cite{Giombi:2009wh,Giombi:2010vg,Giombi:2011kc} for some examples and \cite{Boulanger:2015ova} for the detailed analysis (even though the conclusions of \cite{Giombi:2009wh,Giombi:2010vg,Giombi:2011kc} were mostly positive at the time). The latter effectively reconstructs the bulk theory by inverting the AdS/CFT correspondence and therefore does not by itself provide an independent definition of the bulk dual. Moreover, its relation to the expected bulk description remains somewhat indirect, and it is not yet clear how to extend the construction to Chern–Simons vector models.}

\begin{wrapfigure}{r}{0.4\textwidth}
    \includegraphics[width=0.9\linewidth]{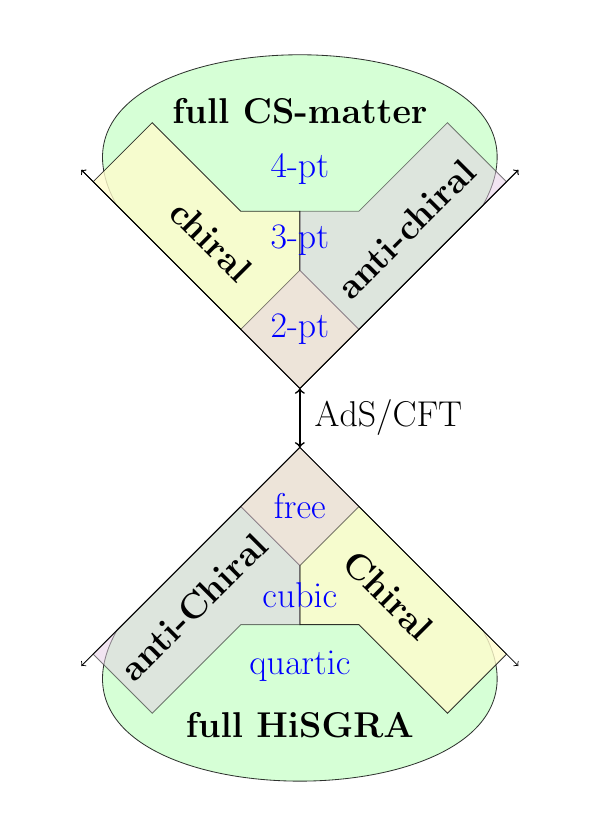}
    \caption{A picture taken from \cite{Jain:2024bza} illustrates the structure of the AdS/CFT duality for Chern--Simons vector models vs. higher-spin gravity. The free spectra coincide in all theories. At the cubic level/three-point functions chiral and anti-chiral interactions provide the complete basis.}
    \label{fig:duality}
\end{wrapfigure}

On the other hand, Chiral HiSGRA is a local (higher-spin) field theory that has the same spectrum as the hypothetical dual of vector models, but necessarily features fewer interactions. Therefore, it has to capture a subsector of Chern--Simons vector models. The basic map of AdS/CFT dualities for the case of Chern--Simons vector models has the structure depicted in Fig. \ref{fig:duality}. In fact, the existence of (anti)-Chiral theories implies that the vector models should have two closed subsectors. These subsectors can be defined via AdS/CFT correlators, but it would be more instructive to have an independent definition within vector models \cite{Jain:2024bza,Aharony:2024nqs}, which seems to require imaginary Chern--Simons level. Let us stress that, being self-dual, Chiral HiSGRA avoids the usual no-go's including the nonlocality arguments both in flat and (A)dS${}_4$ spaces. 

In the duality between Chiral HiSGRA and a subsector of vector models it seems plausible to get the complete solution on both sides. In the paper we take a step toward AdS/CFT for Chiral HiSGRA and, more generally, for all self-dual theories with massless (higher-spin) fields. Following the recent work \cite{Skvortsov:2026gtq} on (SD)YM, we derive the Fefferman-Graham expansion for free self-dual theories, propose a holographic dictionary and derive bulk-to-bulk and boundary-to-bulk propagators for fields of arbitrary spin.   

As a concrete application of these results we consider HS-SDYM, which is a truncation of Chiral theory. This theory (and also HS-SDGR) has some crucial advantages: these theories have simple actions and do not require the advanced machinery of strong-homotopy algebras \cite{Sharapov:2022awp, Sharapov:2022wpz} to be defined. Therefore, we can directly compute AdS/CFT correlators in HS-SDYM, which also gives a small part of Chiral HiSGRA correlators.

The paper is organized as follows. In Section \ref{sec:HSSDYM} we review free self-dual theories and HS-SDYM. In Section \ref{sec:adscft} we derive the Fefferman-Graham expansion and discuss the AdS/CFT dictionary. In Section \ref{sec:correlators} the propagators are derived and three- and four-point AdS/CFT correlators are computed in HS-SDYM. We end up with some conclusions and discussion in Section \ref{sec:conclusions}. A couple of technical Appendices are referred to in due time.

\section{Self-dual (higher-spin) fields}
\label{sec:HSSDYM}
Since HS-SDYM is a straightforward extension of SDYM, let us review the Chalmers--Siegel approach to the latter first. Next, we recall the chiral description of free higher-spin fields, after which HS-SDYM is about putting both ingredients together. 

\paragraph{Chalmers--Siegel action for SDYM.} We begin with the Yang-Mills action (the trace over the color is implied)\footnote{Throughout the paper we use the two-component spinor language: $A,B,...=1,2$, $A',B',...=1,2$ and $\epsilon_{AB}=-\epsilon_{BA}$, $\epsilon_{A'B'}=-\epsilon_{B'A'}$ are the invariant tensors. }
\begin{align}
    S_{\text{YM}}&=\int F_{\mu\nu}^2 =\tfrac12 \int \Big(F_{AB}^2 +F_{A'B'}^2\Big)\,,
\end{align}
where the decomposition into (anti)-selfdual components reads
\begin{equation}
F_{\mu\nu} \equiv F_{AA'BB'} = \tfrac{1}{2}F_{AB}\epsilon_{A'B'} + \tfrac{1}{2}F_{A'B'}\epsilon_{AB} \,.
\end{equation}
By adding the right amount of the topological invariant, the theta-term,
\begin{align}
    \int F\wedge F\sim \int \Big(F_{AB}^2 -F_{A'B'}^2\Big)  
\end{align}
one brings the action to \cite{Chalmers:1996rq}
\begin{align} \label{fullYM}
    \tfrac12 \int F_{AB}^2 \sim \int \Psi^{AB} F_{AB} -\tfrac12 \Psi_{AB}\, \Psi^{AB}\,,
\end{align}
where in the last step we introduced an auxiliary field $\Psi^{AB}$. By dropping the last term, both being gauge invariant, we land on the Chalmers-Siegel action for SDYM
\begin{align}
    S_{SDYM}[A,\Psi]&= \int \Psi^{AB} H_{AB} \wedge F\,,
\end{align}
where $F$ is a two-form field-strength $F=dA+AA$ and we introduced the basis of self-dual two-forms $H^{AB}=e\fud{A}{C} \wedge e^{CB}$ with $e^{AA'}$ being the background vierbein. Here, $F=F(A)$ contributes effectively through its self-dual projection $\tfrac12 F_{AB} H^{AB}$ since $H_{AB} \wedge \bar{H}_{A'B'}\equiv0$.   

\paragraph{Free higher-spin fields.} Massless higher-spin states are often realized by Fronsdal fields \cite{Fronsdal:1978rb}, which are symmetric tensors $\Phi_{\mu_1...\mu_s}$. However, Fronsdal's description has certain limitations, see, e.g., a discussion at the end of \cite{Krasnov:2021nsq}. Another formulation (two, in fact), which is tied to $4d$, originates from twistor space. The two helicities $\pm s$ can be described by a pair of equations proposed by Penrose \cite{Penrose:1965am}\footnote{As we move to higher-spins, a condensed notation becomes useful: a group of symmetric indices $A_1...A_k$ is abbreviated to $A(k)$. Also, indices that need to be symmetrized are denoted by the same letter, e.g. $\xi^A \eta^A\equiv \xi^{(A_1}\eta^{A_2)}=\tfrac12(\xi^{A_1}\eta^{A_2}+\xi^{A_2}\eta^{A_1}) $. }
\begin{align} \label{PsiEOM}
    {\nabla^\ga}_\gbd\Psi^{\gbd\gad\left(2s-1\right)}=0\,,& &{\nabla_\gb}^\gad\Psi^{\gb\ga\left(2s-1\right)}=0\,.
\end{align}
For $s=1$ the equations are just the Maxwell equations (together with the Bianchi identities) for the (anti)-self-dual components $F_{AB}$, $F_{A'B'}$ of the field strength $F_{\mu\nu}$. For $s=2$ the $\Psi$s are the (anti)-self-dual components of the (linearized) Weyl tensor. The equations suffice to describe free fields. However, in order to construct interesting interactions, one of these ``field-strengths'' should be replaced by a gauge potential.\footnote{See \cite{Guarini:2026vds} for a classification of cubic interactions within the chiral approach, which also shows that \eqref{PsiEOM} captures a very limited set of interactions.} As a result, twistors and the spacetime equivalents thereof (after the Penrose transform) treat positive and negative helicities differently. We can refer to these field variables as the chiral ones. This does not imply that the theories are necessarily chiral/self-dual, cf. the Chalmers--Siegel action for Yang-Mills theory here-above. Parity-invariant theories can be reformulated in terms of chiral fields, which may also have certain advantages. 

Let us define $\Psi^{A(2s)}$ to have helicity $-s$ and $\Psi^{A'(2s)}$ to have helicity $+s$. The gauge potential for the $+s$ helicity is $\Phi^{\ga\left(2s-1\right),\gad}$ and it obeys \cite{Hughston:1979tq,Eastwood:1981jy,Woodhouse:1985id}
\begin{align} \label{PhiEOM1}
{\nabla^\ga}_\gad\Phi^{\ga\left(2s-1\right),\gad}=0\,,& &\delta\Phi^{\ga\left(2s-1\right),\gad}=\nabla^{\ga\gad}\xi^{\ga\left(2s-2\right)}\,,
\end{align}
where $\Phi$ can be connected to $\Psi$ through $\Psi^{\gad\left(2s\right)}\sim{\nabla_\ga}^\gad\dots{\nabla_\ga}^\gad\Phi^{\ga\left(2s-1\right),\gad}$. In fact, along the way from $\Phi^{A(2s-1),A'}$ to $\Psi^{A'(2s)}$ one also finds
\begin{align}
    \Phi^{A(s),A'(s)}&= \nabla\fdu{B}{A'}...\nabla\fdu{B}{A'} \Phi^{A(s)B(s-1),A'} \,,
\end{align}
which can be identified with the traceless part of the Fronsdal field, if need be. For $s=1$, $\Phi^{A,A'}$ is just the Maxwell gauge potential $A_\mu$ rewritten in the spinorial language, $\Phi_{A,A'}=\tfrac12 A_\mu\sigma^\mu_{AA'}$. For $s=2$, $\Phi^{AAA,A'}$ is a component of the self-dual part $\omega^{AB}$ of the spin-connection $\omega^{a,b}_\mu \, dx^\mu$. 

Since $\Phi^{A(2s-1),A'}$ is a gauge field, a further improvement of the approach is to introduce a one-form ``connection'' $\omega^{A(2s-2)}$ that contains $\Phi^{A(2s-1),A'}$
\begin{align}\label{eq:omega_def}
    \omega^{\ga\left(2s-2\right)}\equiv e_{\gb\gbd}\Phi^{\ga\left(2s-2\right)\gb,\gbd}+{e^\ga}_\gbd\Theta^{\ga\left(2s-3\right),\gbd}\,,
\end{align}
where $e^{\ga\gad}\equiv e^{\ga\gad}_\mu dx^\mu$ is the background vierbein. In order to remove the extra redundant component one employs a shift symmetry with parameter $\eta^{\ga\left(2s-3\right),\gbd}$
\begin{align}\label{eq:omega_gauge}
    \delta\omega^{\ga\left(2s-2\right)}=\nabla\xi^{\ga\left(2s-2\right)}+{e^\ga}_\gbd\eta^{\ga\left(2s-3\right),\gbd}\,,
\end{align}
where $\nabla=d+...$ is the covariant derivative and $d$ is the exterior derivative. There is a simple gauge-invariant action that replaces the Fronsdal one \cite{Krasnov:2021nsq}
\begin{align} \label{simpleaction}
    S_s= \int\Psi^{\ga\left(2s\right)}\wedge H_{\ga\ga}\wedge\nabla\omega_{\ga\left(2s-2\right)}\,.
\end{align}
As different from the Fronsdal one, the action is gauge-invariant on any self-dual background (not just the maximally symmetric spaces). A self-dual background is such that $\nabla^2 \chi^A\equiv0$ for any $\chi^A$.

\paragraph{HS-SDYM.} To construct a higher-spin extension of SDYM  it is convenient to introduce generating functions. For example, for all positive helicity fields we define
\begin{align}
    \omega\equiv \omega(y|x)&= \sum_s \tfrac{1}{(2s-2)!}\omega_{A_1...A_{2s-2} }\, y^{A_1}...\, y^{A_{2s-2}} \,,
\end{align} 
where $y^A$ is a commuting auxiliary variable. These fields are also assumed to take values in some representation, say, the adjoint, of a Lie algebra. Likewise, we build a generating function $\Psi=\Psi(y|x)$ of negative helicity fields that takes values in the dual module
\begin{align}
    \Psi\equiv \Psi(y|x)&= \sum_{s>0} \tfrac{1}{(2s)!}\Psi_{A_1...A_{2s} }\, y^{A_1}...\, y^{A_{2s}} \,.
\end{align}
Note that we dropped the scalar field from $\Psi$. For two generating functions $F$ and $G$ such that their form degree adds up to $d$ we define the pairing
\begin{align}\label{scalarp}
    \action{F}{G}&= \int \scalar{F}{G}\,, &
    \scalar{F}{G}&= \sum_n \frac{1}{n!} F_{A(n)} \wedge G^{A(n)}\,.
\end{align}
A practically useful way to realize the pairing is $\scalar{F}{G}=F(\pl)G(y)$. The action for SDYM can be rewritten as
\begin{align}   \label{SDYMAction}
  S_{\text{SDYM}}&= \action{\tfrac12\Psi_{AA}y^Ay^A}{\tfrac12 H_{AA}y^Ay^A \wedge F}\,.
\end{align}
Likewise, the sum of the free actions of all $s>0$ fields is
\begin{align} 
  S_{\text{free}}&= \action{\Psi(y)}{\tfrac12 H_{AA}y^Ay^A \wedge \nabla \omega(y)}\,.
\end{align}
This is the same action as \eqref{simpleaction}, but rewritten in the language of generating functions, which is more convenient for a higher-spin generalization.

In order to build gauge interactions we need a non-abelian field-strength $F=\nabla\omega-\tfrac12[\omega\wedge, \omega]$, where the Lie bracket $[\bullet,\bullet]$ leaves $y^A$ intact, i.e. in components we have 
\begin{align}\label{omega^2}
    \tfrac12 [\omega \wedge, \omega] &= \sum_{n,m=0} \frac{1}{2\,n!m!}\,[\omega_{A(n)}, \omega_{A(m)}]\, y^{A_1}...\,y^{A_{n+m}}\,.
\end{align}
The gauge algebra is the loop algebra $\mathbb{C}[y]\otimes \mathfrak{g}$, where $\mathfrak{g}$ is any (color) Lie algebra.\footnote{$\mathfrak{g}$ does not have to have a nondegenerate bilinear invariant form since $\Psi$ takes values in the dual module $\mathfrak{g}^*$ and this pairing is canonical and invariant. } 
The action reads
\begin{align} \label{HS-SDYMaction}
  S_{\text{HS-SDYM}}&= \action{\Psi}{\tfrac12 H_{AA}y^Ay^A \wedge F}\,.
\end{align}
It is invariant under the nonabelian gauge transformation
\begin{align}
    \delta \omega&= \nabla\xi-[\omega,\xi]\,,& 
    \delta \Psi&= [\Psi,\xi]\,,
\end{align}
as well as it is still invariant under 
\begin{align}\label{symmetry-shifts}
    \delta \omega^{A(k)}&= e\fud{A}{C'} \eta^{A(k-1),C'}\,,
\end{align}
which is a simple consequence of $H_{AA}\wedge e_{AB'}\equiv0$. Upon a close examination one sees that action \eqref{HS-SDYMaction} contains only ``gauge''/one-derivative interactions of the Yang-Mills type, which is obvious in the light-cone gauge. It contains vertices of $V_{-++}$-type with $\lambda_1+\lambda_2+\lambda_3=1$ and $\lambda_1<0$, $\lambda_{2,3}>0$. Such interactions are also present in Chiral HiSGRA and all amplitudes of HS-SDYM are subsets of those of Chiral theory.

\section{Self-dual AdS/CFT}
\label{sec:adscft}

In order to make a sensible proposal for self-dual AdS/CFT correspondence we first work out the Fefferman-Graham expansion of free chiral fields. The asymptotic analysis reveals the boundary data that is dual to CFT operators. Next, for the physical degrees of freedom the requirement to have regular solutions in the bulk unambiguously fixes boundary conditions. Lastly, (higher-spin) currents of the dual CFT can be assembled from positive and negative parts provided by the bulk fields.

Since twistors/chiral description discriminate degrees of freedom by helicity and the helicity is a momentum space concept, all of the discussion below is in the hybrid $3+1$ momentum+radial coordinate approach to AdS/CFT. A field is a function $f(\vec k,z)$ of the $3d$ momentum along the boundary and of the radial coordinate $z$ in Poincare coordinates.  
Following \cite{Maldacena:2011nz}, given a $3d$ momentum $k^i\equiv k^{AB}$, let us introduce an auxiliary $4d$ light-like momentum $p_{AA'}=k_{AA'}+|k|\epsilon_{AA'}$.\footnote{Our convention is that $p_\mu p^\mu=-\tfrac12 p_{AA'}p^{AA'}$, $|k|^2=k_i k^i= -\tfrac12 k_{AB}k^{AB}$. } Since it is light-like it can be factorized as $p_{AA'}=k_A\bar{k}_{A'}$. This allows one to define two $3d$ vectors $k_A k_B$ and $\bar{k}_{A}\bar{k}_{B}$ that are $3d$-null and orthogonal to $k^{AB}$. Such auxiliary light-like momenta are very helpful in defining the spinor-helicity formalism for $3d$ CFTs and $\text{(A)dS}_4$, see e.g. \cite{Maldacena:2011nz,Jain:2021vrv}.  

\subsection{Fefferman-Graham expansion}
\label{sec:FG}

We begin with the negative helicity field $\Psi^{A(2s)}$ as it is not a gauge field. Next, we proceed to the positive helicity gauge field $\Phi^{A(2s-1),A'}$. Lastly, we discuss a second order formulation that unifies the previous two. 

\subsubsection{Negative helicity}
The equation of motion for the zero-form field $\Psi^{A(2s)}$ reads
\begin{align} \label{eom}
    {\nabla_\gb}^\gad\Psi^{\ga\left(2s-1\right)\gb}=0\,,
\end{align}
and after applying the $\text{AdS}_4$ covariant derivative in Poincare coordinates one finds
\begin{align}
    (z k\fdu{B}{B'}+\epsilon\fdu{B}{B'}(s+1-z\partial_z))\Psi^{A(2s-1)B} = 0 \,.
\end{align}
We introduce the $z$-expansion as
\begin{align} \label{expans}
    \Psi^{\ga\left(2s\right)}&=z^\Delta \psi^{A(2s)}(z) = z^\Delta \Big(\psi^{A(2s)}_{(0)} +\psi^{A(2s)}_{(1)}z+\psi^{A(2s)}_{(2)}z^2+\dots\Big)\,.
\end{align}
In terms of the field $\psi(z)$ the equation of motion \eqref{eom} reads 
\begin{align}
    \Big(s+1-\Delta\Big)\psi^{A(2s-1)B} + z\Big(k\fdu{B}{B'}-\epsilon\fdu{B}{B'}\partial_z\Big)\psi^{A(2s-1)B} = 0\,.
\end{align}
The first term yields the conformal dimension $\Delta=s+1$. 
The conformal dimension of a spin-$s$ primary operator at the unitarity bound in $d$ dimensions is $\Delta=d-2+s$,
which matches the result, since $d=3$ in our case. Such an operator $J_{a_1...a_s}$ is a conserved higher-spin current, $\pl^m J_{ma_2...a_s}=0$. After imposing this value for the conformal dimension we are left with
\begin{align} \label{nextOrder}
    (k\fdu{B}{B'}-\epsilon\fdu{B}{B'}\partial_z)\psi^{A(2s-1)B}=0\,.
\end{align}
The latter equation has the same form as in flat space thanks to the conformal invariance of \eqref{eom}. 
The free index $B'$ can be either symmetrized or anti-symmetrized with the other free indices. The components of $\psi^{A(2s)}$ satisfy in the former case
\begin{align} \label{rec}
    \psi_{(n)}^{A(2s)}=\frac{1}{n}k\fdu{B}{A}\psi_{(n-1)}^{A(2s-1)B} \,,
\end{align}
while in the latter we find the conservation law in momentum space
\begin{align} \label{cons}
    k_{BB}\psi_{(n)}^{A(2s-2)BB}=0 \,.
\end{align}
Together with the conformal dimension, this confirms that the Fefferman-Graham boundary data of $\Psi^{A(2s)}$ is a totally symmetric conserved spin-$s$ current $J_{a_1...a_s}\equiv J_{A(2s)}$ on the boundary.

Since the recurrence relation \eqref{rec} does not terminate at any order and its form is simple enough, it may be used to construct a solution that extends to everywhere in the bulk. Indeed, the conservation equation leaves us with only two components in $\psi_{A(2s)}$, which are along $k_A...k_A$ and $\brk_A...\brk_A$. The corresponding two solutions are  
\begin{align} \label{PsiProp}
    \Psi^{A(2s)} &= \psi_{(0)}^-z^{s+1}e^{-|k|z}k^{A(2s)}\,, & 
    \Psi^{A(2s)} &= \psi_{(0)}^+z^{s+1}e^{+|k|z}\brk^{A(2s)}\,,
\end{align}
where $\psi_{(0)}^{\pm}$ are the true (unconstrained) Fefferman-Graham boundary data. What we would like to stress is that, similarly to SDYM, whether a solution is regular or not in the bulk (i.e. $e^{-kz}$ or $e^{+kz}$) is tied to the polarization/helicity it has.

\subsubsection{Positive helicity}
Next, we derive the Fefferman-Graham expansion for the higher-spin gauge field $\Phi_{A(2s-1),A'}$, for which we impose the gauge condition $\Phi\fud{B'}{A(2s-2),B'}=0$ (this an analog of the radial gauge $n\cdot A=0$, if $n_{AA'}=\epsilon_{AA'}$, which for $s=1$ is $A_{z}=0$). As a result, the field is totally symmetric and from now on we can also write $\Phi_{A(2s)} \equiv \Phi_{A(2s-1),A}$. The equation of motion \eqref{PhiEOM1} gives
\begin{align}
    \nabla\fdu{A}{B'}\Phi_{A(2s-1),B'} = \Big( zk\fdu{A}{B'} + (-s+2-z\partial_z)\epsilon\fdu{A}{B'} \Big) \Phi_{A(2s-1),B'} = 0\,.
\end{align}
The Fefferman-Graham expansion reads
\begin{align}
    \Phi_{A(2s)} = z^{\Delta}\phi_{A(2s)}(z) = z^\Delta \Big( \phi^{(0)}_{A(2s)} + \phi^{(1)}_{A(2s)} z +\phi^{(2)}_{A(2s)} z^2 + \dots \Big)\,.
\end{align}
In terms of the field $\phi_{A(2s)}$ the equation of motion is rewritten as
\begin{align}
    (-s+2-\Delta)\phi_{A(2s)} + z (k\fdu{A}{B}-\epsilon\fdu{A}{B}\partial_z)\phi_{A(2s-1)B} = 0 \,.
\end{align}
Setting $z=0$ yields the conformal dimension $\Delta=2-s$, 
while the solution can be obtained from
\begin{align}
    (k\fdu{A}{B} - \epsilon\fdu{A}{B}\partial_z)\phi_{A(2s-1)B} = 0\,.
\end{align}
This is the same as \eqref{PhiEOM1}, but in flat space, thanks to the conformal invariance. The conformal dimension $\Delta=2-s$ is the right one for a source $\varphi^{a(s)}$ of a higher-spin current $J_{a_1...a_s}$. For the Taylor coefficients we find
\begin{align}
    \phi^{(n)}_{A(2s)}=\frac{1}{n}k\fdu{A}{B}\phi^{(n-1)}_{A(2s-1)B} \,,
\end{align}
just like \eqref{rec}. Note that $\phi^{(2s-1)}$ is a gauge-invariant ``Cotton'' tensor. Previously we contracted two free indices to obtain a conservation law, see \eqref{cons}, which is equivalent to evaluating the part of the equation that is anti-symmetric in them. Here, such a conservation law cannot be obtained. Indeed, the equation for $\Phi_{A(2s-1),A'}$ is obtained by varying with respect to $\Psi^{A(2s)}$ and vice versa. Therefore, there are ``more'' equations for $\Psi$ than for $\Phi$.

The boundary data $\phi^{(0)}_{A(2s)}$ is a gauge field and modulo the gauge symmetry has $2$ off-shell degrees of freedom, which can be chosen along $k_{A(2s)}$ and $\brk_{A(2s)}$. Alternatively, one can consider the physical gauge where $\Phi_{A(2s-1),A'}$ is both $\epsilon$-traceless, i.e. it is effectively symmetric $\Phi_{A(2s)}$, and is also $k$-transverse.\footnote{In the spin-one case, this corresponds to imposing the Coulomb gauge $k\cdot A=0$ and then integrating out $A_z$, which leads to an instantaneous term in the action (it contributes starting from the quartic order and, hence, can be ignored in the free theory).} This leaves us with the same two directions $k_{A(2s)}$ and $\brk_{A(2s)}$. The corresponding solutions are
\begin{align} \label{PhiProp}
        \Phi_{A(2s)}&=\phi^{(0)}_-z^{2-s}e^{+|k|z} k_{A(2s)}\,, &
        \Phi_{A(2s)}&=\phi^{(0)}_+z^{2-s}e^{-|k|z} \brk_{A(2s)}\,.
\end{align}
We see that in the first order formulation only the latter yields a regular solution. Hence, within AdS/CFT $\Phi$ and $\Psi$ each describe one degree of freedom, the one for which a regular solution exists.

In order to ensure that the field $\Phi_{A(2s-1),A'}$ remains in the same gauge, its gauge parameter should satisfy the condition
\begin{align}
    \delta \Phi\fdu{A(2s-2)B,}{B} = \frac{1}{2s-1} \Big( \nabla\fdu{B}{B}\xi_{A(2s-2)}+(2s-2)\nabla\fdu{A}{B}\xi_{A(2s-3)B} \Big) = 0\,.
\end{align}
Explicitly, this yields the equation
\begin{align} \label{xiEq}
    \Big( (2s-2)zk\fdu{A}{B} + 2s(-s+1-z\partial_z)\epsilon\fdu{A}{B} \Big) \xi_{A(2s-3)B} = 0 \,.
\end{align}
We write the expansion
\begin{align}
    \xi_{A(2s-2)} = z^\Delta\eta_{A(2s-2)}(z)= z^{\Delta} (\eta^{(0)}_{A(2s-2)}+\eta^{(1)}_{A(2s-2)}z+\eta^{(2)}_{A(2s-2)}z^2+\dots)
\end{align}
and the equation for $\eta_{A(2s-2)}$ reads
\begin{align}
    \Big( 2s(-s+1-\Delta)\epsilon\fdu{A}{B}+z\big( (2s-2)k\fdu{A}{B}-2s\epsilon\fdu{A}{B}\partial_z \big) \Big) \eta_{A(2s-3)B} = 0\,.
\end{align}
Setting $z=0$ yields the conformal dimension $\Delta = 1-s $, 
which is the right value for the gauge parameter of the source $\varphi^{a(s)}$ of a conserved spin-$s$ current, while the solution can be found from (the equation for the gauge parameter is not conformally invariant)
\begin{align}
    \Big( (2s-2)k\fdu{A}{B}-2s\epsilon\fdu{A}{B}\partial_z \Big)\eta_{A(2s-3)B} =0 \,.
\end{align}
The coefficients in the expansion are expressed by
\begin{align}\label{eta}
    \eta^{(n)}_{A(2s-2)} = \frac{1}{n}\frac{2s-2}{2s}k\fdu{A}{B}\eta^{(n-1)}_{A(2s-3)B}\,.
\end{align}
Note that the equation only holds when all free indices are symmetrized. Hence, no conservation law as in \eqref{cons} can be extracted.
The case when $s=1$ is special, as \eqref{eta} shows that all higher coefficients are zero, except for $\eta^{(0)}$. This can be understood from the fact that for $s=1$ \eqref{xiEq} simply requires $\partial_z\xi=0$. Note that the first  coefficient $\phi^{(1)}_{AA}=k\fdu{A}{B} \phi^{(0)}_{AB}$ is gauge invariant under $\delta \phi^{(0)}_{AA}=k_{AA} \eta^{(0)}$. It is just the field strength of the boundary gauge field $\phi^{(0)}_{AA}$. The gauge variation of $\Phi_{A(2s)}$ is then
\begin{align}
    \delta\Phi_{A(2s)} = \nabla_{AA}\xi_{A(2s-2)} = zk_{AA}\xi_{A(2s-2)} \,,
\end{align}
which has the expansion
\begin{align}
    \delta\Phi_{A(2s)} = z^{2-s}k_{AA}\Big( \eta^{(0)}_{A(2s-2)} + z \eta^{(1)}_{A(2s-2)} + \dots \Big)
\end{align}
and we see that it has the right scaling behavior at the boundary. Moreover, the boundary data $\phi^{(0)}_{A(2s)}$ transforms as $\delta\phi^{(0)}_{A(2s)} = k_{AA}\eta^{(0)}_{A(2s-2)}$.

\subsubsection{Second order formulation}
Let us recall that the Fefferman-Graham expansion of the Fronsdal field has two boundary data \cite{Mikhailov:2002bp,Bekaert:2012vt}: a gauge field $\phi_{A(2s)}$ and a conserved current $J_{A(2s)}$:
\begin{align}
    \Phi_{A(s),A'(s)}&= z^{2-s} \phi_{A(s)A'(s)} +... + z^{s+1} J_{A(s)A'(s)} +... \,.
\end{align}
Here, we consider only the traceless part $\Phi_{A(s),A'(s)}$ of the Fronsdal field $\Phi_{\mu_1...\mu_s}$. Except for $s=1$ where $\Phi_{A,A'}$ is shared by both the first order chiral and by the second order (Maxwell) formulations, the first order chiral and the second order Fronsdal formulations are quite different. Nevertheless, there is another second order formulation that is closely related to the chiral one. 

The main idea is the same as for the Chalmers--Siegel action: the $\Psi^2$-term can be added to SDYM action to get Yang-Mills theory. With the help of the Chalmers-Siegel trick we get
\begin{align}\label{epsilonaction}
    S = \int |e| \Psi^{A(2s)}\nabla_{AB'}\Phi\fdu{A(2s-1),}{B'}-\frac{\epsilon}{2}\int |e| \Psi^{A(2s)}\Psi_{A(2s)}\,.
\end{align}
For $\epsilon\neq0$ the resulting second order action in terms of the physical field $\Phi_{A(2s-1),A'}$, after integrating out $\Psi$, gives effectively
\begin{align}
    S&= \int |e|\, \left(\nabla\fdu{A}{B'}\Phi_{A(2s-1),B'}\right) \left(\nabla^{AC'}\Phi\fud{A(2s-1),}{C'}\right)\,.
\end{align}
It is easy to show (in flat space, for example) that the action describes helicity $\pm s$ states. Evidently, a natural second order formulation that is related to the chiral formulation is not the Fronsdal one. A parameter $\epsilon$ may be introduced that controls whether the formulation is first or second order. In the limit $\epsilon \rightarrow 0$, one finds the first-order formulation. However, the limit is not smooth since $\epsilon\neq0$ and $\epsilon=0$ formulations are different, but describe the same number of degrees of freedom. 

Without going into much detail, let us state the main results of the Fefferman-Graham analysis of the second order formulation. The asymptotic expansion combines the two of the first order chiral descriptions
\begin{align}
    \Phi_{A(2s-1),A'}&= z^{2-s} \phi_{A(2s-1)A'} +... + z^{s+1} J_{A(2s-1)A'} +...\,.
\end{align}
The leading falloff is a gauge field on the boundary $\delta \phi_{A(2s)}=k_{AA}\eta_{A(2s-2)}$ and the subleading falloff is a conserved spin-$s$ current $J_{A(2s)}$.

\subsection{Boundary conditions, AdS/CFT dictionary}
\label{sec:}
Admissible boundary conditions correspond to a Lagrangian submanifold of the Fefferman-Graham data that is compatible with regularity in the bulk. It is easier to analyze the problem in the physical gauge, where both $\Psi_{A(2s)}$ and $\Phi_{A(2s-1)A'}\equiv \Phi_{A(2s-1),A'}$ have two off-shell physical degrees of freedom. Let us introduce two polarization spinors ($\epsilon_\pm \cdot \epsilon_\pm=0$, $\epsilon_-^A \epsilon_+^B\epsilon_{AB}=1$)
\begin{align}\label{polarizationspinors}
    \epsilon_+^A(\vec k)&=  \frac{\brk^{A}}{\sqrt{2k}}\,, & \epsilon_-^A(\vec k)&=  \frac{k^{A}}{\sqrt{2k}}\,,
\end{align}
and decompose $\Phi$ and $\Psi$ as
\begin{align}
\begin{aligned}
    f^{A(2s)}(\vec k,z)&= \epsilon_+^{A}...\epsilon_+^{A}f_+(\vec k,z)+\epsilon_-^{A}...\epsilon_-^{A}f_-(\vec k,z)\,,
\end{aligned}
\end{align}
where $f$ is either $\Psi$ or $\Phi$. The free action reduces to 
\begin{align}
    S&= \int_M (\Psi_+, D_+ \Phi_+)+(\Psi_-, D_- \Phi_-)\,,
\end{align}
where $(f,g)\equiv f(-k,z) g(+k,z) $ and $D_\pm=\pl_z \pm k$. It is clear that $D_+f=0$ has only regular solutions $\sim e^{-kz}$, while $D_-f=0$ has only irregular solutions $\sim e^{+kz}$. This is consistent with and is also a compact way to summarize the Fefferman-Graham analysis of the previous section. The irregular modes must be frozen.\footnote{These boundary conditions can also be called APS (Atiyah--Patody--Singer) since the idea is the same: eliminate negative eigenvalues in the radial direction and the kinetic operators are extensions of the Dirac operator to higher spins. } Since the equations are first order, the only boundary problem we can afford is the Dirichlet one. However, we have to set $\Psi_+=0$, $\Phi_-=0$ at the boundary to respect the regularity, while $\Phi_+$, $\Psi_-$ are free at the boundary. The action is clearly stationary with such boundary conditions.\footnote{Let us note that one can add contact boundary terms of type $\Psi\Phi$ for various helicity configurations, if needed. Any first order action without a boundary term leads to vanishing two-point function as the on-shell action vanishes. }

Let us recall that both the chiral and the Fronsdal second order formulations have the same boundary data: a gauge field $\phi_{A(2s)}$ and a conserved higher-spin current $J_{A(2s)}$. The standard Dirichlet holography fixes the gauge field on the boundary $\phi$. Then, the on-shell action gives a generating functional $W_{D}[\phi]$ of correlators $\langle j...j\rangle $ of conserved (higher-spin) currents on the boundary (we denote them $j$ to distinguish from the boundary data $J$). Another scenario is to fix $J$ on the boundary, i.e. Neumann problem, the on-shell action gives a generating functional $W_N[h]$ of correlators of (higher-spin) gauge fields $h\equiv h_{A(2s)}$ on the boundary \cite{Giombi:2013yva}. At the free level, for each $h$ one can construct a gauge-invariant ``Cotton'' tensor 
\begin{align}
    C^{A(2s)}&= \pl\fud{A}{B}...\pl\fud{A}{B} h^{B(2s-1)A}\,.
\end{align}
The Cotton tensor obeys the Bianchi identity, $\pl_{BB}C^{A(2s-2)BB}=0$, which makes it effectively look like a conserved higher-spin current. 
For $s=1$, $C^{AA}$ is dual to the field strength $F=dA$ and is the well-known dual current $j=*dA$ \cite{Witten:2003ya}. For $s=2$, $C^{A(4)}$ is the linearized Cotton tensor. Therefore, Cotton tensors play the role of dual (higher-spin) currents for all spins. Passing to the dual higher-spin currents makes the resulting CFT be of the same type as the Dirichlet CFT, i.e. a theory with higher-spin currents. As is well-known \cite{Witten:2001ua,Klebanov:1999tb,Hartman:2006dy,Giombi:2011ya}, there is a close relation between the Dirichlet and Neumann CFT duals. In the large-$N$ limit the Neumann CFT is a double-trace deformation of the Dirichlet one
\begin{align}
    S_{\text{CFT}}^{\text{N}}=S_{\text{CFT}}^{\text{D}}+\int j^2 = S_{\text{CFT}}^{\text{D}}+\int j \cdot h -h^2/2 \sim S_{\text{CFT}}^{\text{D}}+\int j \cdot h\,,
\end{align}
where the last equality is true in the large-$N$ limit. Upon expanding $\langle \exp \int j \cdot h\rangle_{\text{CFT}}$ the kinetic term for the gauge field $h_{A(2s)}$ is the two-point function $\langle j_{A(2s)}j_{B(2s)}\rangle$ of the (higher-spin) currents in the Dirichlet CFT. 

Recall that both a conserved higher-spin current $J_{A(2s)}$ and a higher-spin gauge field $\phi_{A(2s)}$ have two off-shell degrees of freedom. With the help of the polarization spinors $\epsilon^{A}_{\pm}(k)$ one can isolate them into $J_\pm$ and $\phi_\pm$. The second order formulations do not discriminate by helicity and we can freely choose to fix either both $\phi_\pm$ or both $J_\pm$ (or even a linear combination of them, i.e. to impose mixed boundary conditions). The first order chiral description does discriminate by helicity and only $\Phi_+$ and $\Psi_-$ can be free at the boundary. If we forget about this feature, then the boundary data of $\Phi$ and $\Psi$ correspond to $\phi$ and $J$:
\begin{align}
\begin{aligned}
    +s&: & \Phi_{A(2s-1),A'}&= z^{2-s} \phi_{A(2s-1)A'} +...\,,\\
    -s&: & \Psi_{A(2s)}&= z^{s+1} J_{A(2s)}+...\,.
\end{aligned}
\end{align}
Thanks to the second order formulation the AdS/CFT meaning of this data is known. What the chiral formulation does is to force to impose ``mixed'' boundary conditions where we fix $\phi_+$ and $J_-$, i.e. a half of the current and a half of the gauge field. The on-shell action should give a functional $W_{SD}[\phi_+,J_-]$ of correlation functions of a half of a current $j_+$ and of a half of the gauge field $h_-$. Via the Cotton tensor/dual higher-spin current we can map\footnote{Note that $k\fud{A}{B} k^B=-k k^A$, $k\fud{A}{B} \brk^B=+k \brk^A$, i.e. $\slashed k$ preserves the helicity and, hence, the Cotton tensor, which is built from $(2s-1)$ factors of $\slashed k$, preserves it as well. } $h_-$ to some $j_-$, thereby, adding the missing half to $j_+$. As a result, we have a CFT with the same spectrum of conserved higher-spin currents as for the Dirichlet and Neumann holography. 

To make the relation between the standard Dirichlet/Neumann/Mixed and self-dual holography more direct, it would help to extend the Chalmers--Siegel and Plebansky/Krasnov ideas from the spin-one and spin-two cases, respectively, to the case of an arbitrary spin-$s$. For example, there should exist a parent action that relates the Fronsdal formulation in terms of symmetric tensors to the second order chiral formulation where the self-dual limit is simple. Similar to the fact that Dirichlet and Neumann results are related via a Legendre transform, i.e. one can get $W_{\text{N}}$ from $W_\text{D}$ without having to recompute the bulk diagrams \cite{Hartman:2006dy,Giombi:2011rz}, we expect that one can also get an access to $W_{\text{SD}}$.

\section{AdS/CFT correlators}
\label{sec:correlators}
Let us illustrate the discussion above and show that one can reliably compute holographic $3$- and, most importantly, $4$-point correlators in (chiral) higher-spin theories. To compute holographic correlators we need to derive bulk-to-bulk propagators, from which the boundary-to-bulk propagators follow as a limit. While in flat space the propagators in the chiral formulation are simple and known for any spin, see e.g. \cite{Guarini:2026vds}, their AdS${}_4$ cousins are not yet available, except for the spin-one case \cite{Skvortsov:2026gtq}. There are two issues to consider: (i) since it is a gauge theory, we need to gauge fix it; (ii) since AdS${}_4$ has a boundary, certain boundary conditions have to be imposed, which is not independent from the gauge fixing. In general, the propagator consists of two parts (omitting the indices for the time being): 
\begin{align}
    \langle \Psi(-k,z) \Phi(k,z')\rangle_{\text{inh}}+\langle \Psi(-k,z) \Phi(k,z')\rangle_{\text{hom}} \,.
\end{align}
The inhomogeneous one is a particular Green's function and the homogeneous one is here to help to impose the required boundary conditions. One can consider different gauges, say $X$ and $Y$. Any two gauges are related via a gauge transformation
\begin{align}\notag
    \langle \Psi^{A(2s)}(-k,z) \Phi^{B(2s-1),B'}(k,z')\rangle_{\text{X.}}&=\langle \Psi^{A(2s)}(-k,z) \Phi^{B(2s-1),B'}(k,z')\rangle_{\text{Y.}}+\nabla^{BB'} \xi^{A(2s),B(2s-2)} \,.
\end{align}
Since the dual theory is a gauge theory one can expect the correlators to be gauge dependent in general. For example, there is a clear gauge dependence in Neumann correlators \cite{Skvortsov:2026gtq} and the self-dual holography seems to be in between Dirichlet and Neumann ones. 

It is easy to see that, similarly to flat space, there is only one family of nonvanishing correlators in (HS)-SDYM, which have the form 
\begin{align}
    \langle \Psi^{A(2s)}(k_1,0) ... \Psi^{C(2s)}(k_{n-1},0) \Phi^{D(2s-1),D'}(k_{n},0)\rangle \,.
\end{align}
In practice, all external lines $\Psi^{A(2s)}$ and $\Phi^{B(2s-1),B'}$ will be replaced by boundary-to-bulk propagators $\langle \Psi^{A(2s)}(-k,0)\Phi^{B(2s-1),B'}(k,z)\rangle $ and $\langle \Psi^{A(2s)}(k,z)\Phi^{B(2s-1),B'}(-k,0)\rangle $, respectively. In addition, the external indices will be contracted with appropriate helicity structures: $\epsilon_-^{A(2s)}$ and $\epsilon_+^{A(2s)}$. Similarly to flat space amplitudes in (HS)-SDYM there is a unique choice of polarizations for which the correlator does not vanish.

As is well-known, the coefficient of the leading energy pole of AdS/CFT correlators should be equal to the amplitude in the same theory but in flat space \cite{Maldacena:2011nz}. The flat space amplitudes vanish in (HS)-SDYM for generic kinematics,\footnote{See, however, the recent \cite{Guevara:2026qzd} and followups \cite{Guevara:2026qwa,Brandhuber:2026njb,Serrani:2026azw}.} and this is one of the simplest tests that four-point functions will have to pass.

\subsection{Propagators}
\label{sec:props}
AdS/CFT bulk-to-bulk and boundary-to-bulk propagators in the chiral formulation have never been considered in the literature for $s>1$ and we first need to fill in this gap.\footnote{There is a vast literature on scalar, vector and tensor propagators and a bit about the arbitrary spin case in the Fronsdal formulation \cite{Bekaert:2014cea}. Propagators for chiral fields in flat space were recently considered in \cite{Guarini:2026vds}.} One very useful observation is that the free theory is conformally invariant. Therefore, instead of AdS${}_4$ space we are going to work with flat space $\mathbb{R}^4$, $z\geq0$ with a boundary at $z=0$. Let us recall the free action for a spin-$s$ particle with the shift symmetry fixed to reduce $\omega^{A(2s-2)}$ to $\Phi^{A(2s-1),A'}$:
\begin{align}
    S_2&=\int \Psi^{A(2s)} \nabla_{AC'}\Phi\fdu{A(2s-1),}{C'} \,.
\end{align}
The free BRST transformations read
\begin{align}
    s\Psi^{A(2s)}&=0\,, & s\Phi^{A(2s-1),A'}&=\nabla^{AA'} c^{A(2s-2)}\,, & s \bar{c}^{A(2s-2)}&=B^{A(2s-2)}\,, & sB^{A(2s-2)}&=0  \,.
\end{align}
In order to gauge fix the theory we can use the same tools as for Yang-Mills theory. One can choose a gauge condition $G(\Phi)$, e.g. $G^{A(2s-2)}=\nabla _{CC'} \Phi^{A(2s-2)C,C'}$ for Feynman/Lorenz gauge and $G^{A(2s-2)}=n _{CC'} \Phi^{A(2s-2)C,C'}$ for an axial gauge (it is convenient to have $n_{CC'}=\epsilon_{CC'}$). Using the standard gauge fixing fermion $\psi=\bar{c}\cdot (B \xi/2 -G)$ we find the ghost and the gauge fixing terms to be (the ghost kinetic term was brought to the canonical form via integration by parts)
\begin{align}
    S_g&=\int s\psi=\int B \cdot (\tfrac{\xi}2 B- G) - \pl_\mu \bar{c} \cdot\pl^\mu c\,.
\end{align}
For the Lorenz/Feynman gauge the BRST variation of the total action $S=S_2+S_g$ vanishes up to the integration by parts that was done to get $\pl_\mu \bar{c} \cdot\pl^\mu c$ instead of $ \bar{c} \cdot\pl_\mu\pl^\mu c$. To be precise, $sS=\int_{\pl M} B\cdot  \pl_z c$ for the Lorenz/Feynman gauge. The variation of the action is (the boundary $z=0$ is the lower limit of integration, hence, the overall minus below) 
\begin{align}
    -\delta S&= \int_{\pl M} \Psi^{A(2s)}\delta \phi_{A(2s)}+B_{A(2s-2)} \delta \rho^{A(2s-2)}-\delta \bar{c}\cdot \pl_zc-\pl_z\bar{c}\cdot\delta c\,.
\end{align}
Above, we introduced the following $3+1$ decomposition of $\Phi^{A(2s-1),A'}$:
\begin{align}
    \Phi^{A(2s-1),B'}&= \phi^{A(2s-1)B'} +\epsilon^{AB'} \rho^{A(2s-2)} && \leftrightarrow && \Phi= (\bry \pl) \phi+ (\bry y) \rho\,.
\end{align}
Here, $(\xi\eta)\equiv \xi^C\eta_C$. The r.h.s. is the same split in terms of generating functions and an auxiliary $\bry^{A'}$ was introduced to hide the primed index of $\Phi$, i.e. all $\Phi$s can be packaged into generating function $\Phi(y,\bry)$ that is linear in $\bry^{A'}$. For $s=1$ the split corresponds to splitting $A_\mu$ into $A_i$ and $\rho=A_z$ since $\epsilon_{AA'}$ is associated to the $z$-direction, i.e. $A^{B,B'}=A^{(B,B')}+ \epsilon^{BB'} \rho$. In terms of the $3+1$ split the gauge symmetries act
\begin{align}
    \delta \phi^{A(2s)}&= k^{AA}\xi^{A(2s-2)} \,,& \delta \rho^{A(2s-2)}&=-\pl_z\xi^{A(2s-2)} \,.
\end{align}
At this point we have to choose a gauge. The axial gauge, $A_z=0$, is perhaps the most convenient one for the AdS/CFT correspondence. It also leads to relatively simple propagators \cite{Raju:2011mp,Raju:2012zs}, including SDYM \cite{Skvortsov:2026gtq}. However, a preliminary analysis shows that propagators in the axial gauge become increasingly more complicated already in flat space as spin grows.\footnote{This is quite clear for $s=2$, see e.g. \cite{Raju:2012zs}.} In this paper we discuss two and half gauges: Feynman/Lorenz gauge and the complete physical gauge. 

\subsubsection{Physical gauge.} 
The complete physical gauge is not strictly speaking a gauge. We first impose the Coulomb gauge $k\cdot \phi=0$. This implies that $\phi$ has only two degrees of freedom, one along $k_A...k_A$ and another along $\brk_A...\brk_A$, which correspond to $\Phi_\pm$ introduced before. Next, we integrate out the longitudinal degrees of freedom, which leads to instantaneous, $1/k^2$, vertices that start at the quartic order. As a result we have an ``effective action'' for only the transverse degrees of freedom: $k\cdot \phi=0$ and $k\cdot \psi=0$, where we introduce the split $\Psi^{A(2s)}=\psi^{A(2s)}+k^{AA}\sigma^{A(2s-2)}$. 

The main advantage of the physical gauge is that the propagator itself is simple. Disadvantages are that there are instantaneous interactions and the final results are more difficult to interpret. In the second order formulation in terms of $\phi$ the flat space propagator is simply
\begin{align}
    \langle \phi^{A(2s)} \phi^{B(2s)}\rangle&= \frac{1}{p^2} \Pi_\text{e}^{A(2s),B(2s)}\,,
\end{align}
where $\Pi_\text{e}$ is the projector onto the $k$-transverse subspace. For $s=1$ it is $\delta_{ij}-k_ik_j/k^2$. For arbitrary $s$ it reads
\begin{align}
    \Pi_\text{e}^{A(2s),B(2s)}=\frac1{(4k^2)^s}\left(\brk^{A(2s)} k^{B(2s)}+k^{A(2s)} \brk^{B(2s)}\right)\,.
\end{align}
The $\langle \psi \phi\rangle$-propagator can be obtained either directly in the first order formulation or via $\psi=\slashed p \phi$ from the second order one ($p^{AA'}=k^{AA'}+\epsilon^{AA'} \omega$): 
\begin{align}
    \langle \psi^{A(2s)} \phi^{B(2s)}\rangle&=p\fud{A}{C'}\langle \phi^{A(2s-1)C'} \phi^{B(2s)}\rangle= \frac{1}{p^2}\left( k \Pi_\text{o}^{A(2s),B(2s)}-\omega \Pi_\text{e}^{A(2s),B(2s)}\right)\,,
\end{align}
where we also introduced the odd projector (for $s=1$ it is $\epsilon^{ijm} k_m/k$)
\begin{align}
    \Pi_\text{o}^{A(2s),B(2s)}=\frac1{(4k^2)^s}\left(\brk^{A(2s)} k^{B(2s)}-k^{A(2s)} \brk^{B(2s)}\right)=\tfrac{1}{k}k\fud{A}{C}\Pi_\text{e}^{A(2s-1)C,B(2s)}\,.
\end{align}
The story above can easily be uplifted to AdS${}_4$ realized as a half of $\mathbb{R}^4$. Let us recall that $D_+D_-=\pl^2_z-k^2$ and the Neumann ($+$) and Dirichlet ($-$) propagators for the conformally-coupled scalar field are
\begin{align}\label{propscalar}
    G&=-\frac{1}{2k}\left(e^{-k|z-z'|}\mp e^{-k(z+z')}\right)=f(z-z')\mp f(z+z') \,.
\end{align}
The first term here is the Fourier transform of $1/p^2=1/(k^2+\omega^2)$. $f$ satisfies
\begin{align}
    \pl_z f&=-k \sign(z-z')f\,,  &\pl^2_zf&=\delta(z-z') +k^2f\,, &&f=-\tfrac{1}{2k}e^{-k|z-z'|}\,.
\end{align}
The second term in \eqref{propscalar}, which is a solution of the homogeneous equation obtained via the method of images, is needed to achieve the required boundary conditions. Likewise, in order to derive the propagator in the first order theory it is convenient to write
\begin{align}
    \langle \phi^{A(2s)} \phi^{B(2s)}\rangle&= \Pi_\text{e}^{A(2s),B(2s)} f(z-z')\,,
\end{align}
and apply the covariant derivative\footnote{The second order theory is not conformally invariant and the Weyl rescaling by $z^\Delta$ of the AdS${}_4$-action will not give the same action on flat space. We mostly refer to the second order theory to have a simpler presentation of the propagators as derivatives of something simple. Indeed, in the second order chiral theory $\Psi$ satisfies the same equation as in the first order theory. }
\begin{align}\notag
    \langle \psi^{A(2s)}(-k,z) \phi^{B(2s)}(k,z')\rangle&=\nabla\fud{A}{C'}\langle \phi^{A(2s-1)C'} \phi^{B(2s)}\rangle= \\ \label{PropPhysical}
    &=\left( -k \Pi_\text{o}^{A(2s),B(2s)}  +\Pi_\text{e}^{A(2s),B(2s)}\pl_z\right)f(z-z')\,.
\end{align}
Equivalently, one can eliminate the indices by projecting onto the physical components
\begin{align} \label{propPhysicalComp}
    \begin{aligned}
       \langle \Psi_+(-k,z)\Phi_+(k,z')\rangle &= \frac12 e^{-k|z-z'|}[\sign(z-z')-1]=-e^{-k|z-z'|} \theta(z'-z)\,,\\
        \langle \Psi_-(-k,z)\Phi_-(k,z')\rangle &=   \frac12 e^{-k|z-z'|}[\sign(z-z')+1]=+e^{-k|z-z'|} \theta(z-z') \,.
    \end{aligned}
\end{align}
There cannot be any homogeneous solutions since $D_-f=0$ has only irregular solutions, which is a nice property of the physical gauge. We only have to deal with longitudinal modes, which is an equivalent of solving Gauss law in the second order formulation. 

\paragraph{Berends--Giele, Gauss law.} We have to be more careful about the coefficients and take interactions into account. Let us recall that $F= d\omega+\omega \omega$ and that the action of HS-SDYM is $ \left \langle \Psi(\pl) | \tfrac12 (yHy) F\right \rangle$. Eliminating via the local shift symmetry the extra components of $\omega$ compared to $\Phi$ we can choose $\omega= (\pl e\plb) \Phi$, i.e. $\omega^{A(2s-2)}=e_{CC'}\Phi^{A(2s-2)C,C'}$, where we introduced $(\xi \pl \eta)\equiv \xi_C \pl^{CC'}\eta_{C'}$. The relevant component of $F$ is along $H_{MM}$ since the action already has $(yHy)$, which projects $\bar{H}_{A'B'}$ out. At the end of the day, only the symmetric component $(yFy)\equiv y_A F^{AB} y_B$ of $F_{AB}$, $F=\tfrac12 F^{AB}H_{AB}+\mathcal{O}(\bar{H})$, is needed
\begin{align}\label{nfactor}
    (yFy)&= (y\nabla \plb) N\Phi+(\plb_{C'} N\Phi) (\plb^{C'}N \Phi)\,.
\end{align}
What appears is $N\Phi$ and since $\Phi$ was defined with a somewhat arbitrary normalization we redefine $N\Phi \rightarrow \Phi$. With the $3+1$ split $\Phi= (\bry \pl) \phi+ (\bry y) \rho$ the curvature decomposes as
\begin{align}
    (yFy)&= (y\nabla \pl) \phi+\pl_{C} \phi \pl^{C} \phi+(yk y)\rho +[N\phi,\rho]\,.
\end{align}
Let us decompose $\phi$ and $\Psi$ further as
\begin{align}
    \Phi&=(\bry \pl)\phi+ (\bry y)\rho\,,  &
    \phi&= q+ (yky) \chi\,, & \psi&=\pi+ (yky)\sigma\,,
\end{align}
where $k\cdot q=k\cdot \pi =0$. We can impose Coulomb gauge $k\cdot \phi=0$, i.e. $\chi=0$. There are two ways to proceed now. The first one is to vary the action with respect to $\rho$ to find a Gauss law of the following schematic form (this is an expansion of $(\pl k \pl) \Psi =[N\phi, \Psi]$)
\begin{align}
    k_{BB} \Psi^{BBA(2s-2)}&= \sum_k [\phi_{B(2k)},\Psi^{A(2s-2)B(2k)}] \,.
\end{align}
This equation can be solved iteratively with respect to $\sigma$. At the free level $\phi=q$, $\Psi=\psi$, i.e. $\sigma=0$ and we have only the physical degrees of freedom. The solution can be plugged back into the action, which generates an instantaneous interaction of type $[\psi,\phi]k^{-2} [\phi,\phi]$ at the quartic order. Alternatively, we can consider the equation with respect to $\Psi$:
\begin{align}\label{BG}
    (y\nabla \pl) \phi+\pl_{C} \phi \pl^{C} \phi+(yk y)\rho +[N\phi,\rho]=0\,.
\end{align}
In SDYM and HS-SDYM the only nonvanishing correlators are of type $\langle \Psi^n \Phi\rangle$. This means that one can solve \eqref{BG} up to order $n$ and in the last step one has to amputate the bulk-to-bulk propagator and contract the result with the plane wave of $\Phi$. This gives the same result as contracting the boundary-to-bulk propagator for $\Psi$ in the last step instead of the bulk-to-bulk propagator. Essentially, this is the Berends--Giele method \cite{Berends:1987me}, the only difference is that the external line at order $n$ needs to be contracted with a boundary-to-bulk propagator, which replaces flat space polarization spin-tensors. 

What makes the Berends--Giele approach very efficient in flat space is a particular choice of the polarization spin-tensors: $\Phi^{A,A'}\sim q^A \brk^{A'}$, where $q^A$ is a reference spinor that is shared by all $\Phi$s. The reference spinor appears as an overall factor at all orders, which simplifies the recursion. This is no longer true in AdS/CFT. We have to choose polarizations that have definite helicity: $\Phi^{A,A'}\sim \brk^A\brk^{A'}$, i.e. $q^A$ depends on the momentum of the corresponding leg. Lastly, note that $[N\phi,\rho]$ vanishes on free fields and, hence, this term does not contribute to four-point correlators. 

\paragraph{Gauss law.} Now we are going to solve the Gauss law, i.e. solve for $\rho$, in two ways: with indices and in terms of generating functions. The Gauss law can be obtained by acting with $(\pl k\pl)$ on \eqref{BG}
\begin{align}\label{rho}
    (\pl k\pl)(y ky) \rho &= (\pl k \pl) J\,, &&J= \pl^C \phi \pl_C\phi\,.
\end{align}
Note that the term $(y\nabla \pl)\phi$ has disappeared thanks to the Coulomb gauge: in the Gauss law there are no time, i.e. $\pl_z$, derivatives. Any spin-tensor can be decomposed in the basis generated by $k^A$ and $\brk^{A}$. We have only symmetric spin-tensors and any generating function $f(y)$ is a function of $(ky)$ and $(\brk y)$ in this basis. In particular, $(\pl k \pl)= (k\pl)(\brk \pl)$ and $(yky)=(ky)(\brk y)$. 

Any generating function of spin-tensors can be rewritten as a function of $y_k\equiv (ky)$ and $y_\brk\equiv (\brk y)$. This can be achieved with the help of the identity operator:\footnote{One starts with a trivial identity $\mathrm{id}=\sum_n\tfrac{1}{n!}y^{A_1}...y^{A_n} \pl_{A_1}...\pl_{A_n}$ and replaces all implicit $\epsilon$s that contract indices with $\epsilon^{AB}=\tfrac{1}{2k}(k^{A}\brk^B- k^B\brk^{A})$. }
\begin{align}
    \mathrm{id}&= :\exp\left[\frac{-(ky)(\brk \pl) +(\brk y)(k\pl)}{2k}\right]: 
\end{align}
Here, the normal ordering symbol $:\bullet:$ means that all $\pl/\pl y$ should be kept to the right of all $y$s. It acts on any function of $y$ and decomposes its Taylor coefficients in the $k,\brk$ basis, e.g. 
\begin{align}
    J(y)=J(y_k,y_\brk)&= \mathrm{id}\, J(y) |0\rangle\,, 
\end{align}
where $|0\rangle $ stresses that $\pl$ differentiates $J(y)$ and sets $y=0$ at the end. If a given function $f(y)$ is already expressed in terms of $y_k=(ky)$ and $y_{\bar{k}}=(\brk y)$, then we have 
\begin{align}
    (\pl k\pl)(y ky) f(y_k,y_{\bar{k}})&= -4k^2 (N_{y_k}+1)(N_{y_\brk}+1) f \,.
\end{align}
Therefore, the solution to the Gauss law reads\footnote{We used the formal inverse $(N_x+1)^{-1} f(x)=\int_0^1 dt\, f(xt)$. }
\begin{align}
    \rho&= -\frac{1}{4k^2} \int_0^1 dt\, dt' \, :\exp\left[\frac{-t(ky)(\brk \pl) +t'(\brk y)(k\pl)}{2k}\right]: (k\pl)(\brk \pl) J(y)|0\rangle \,.
\end{align}
After Taylor expanding the expression above we find (note that $f(y)= \sum_k f_k y^k/k!$ and $k!$ matters when expanding in order to convert a generating function into explicit index form) 
\begin{align}
    \sum _{m,n} C^{m+n+2}_{m+1} (-) \frac{(-)^m k^{A(m)} \brk^{A(n)} \brk^{B(m+1)} k^{B(n+1)} }{(2k)^{m+1}(2k)^{n+1}} J_{B(m+n+2)}\,.
\end{align}
The same can be achieved in the index form, which is more transparent. We decompose $\rho$ in the basis spanned by $(yk)$ and $(y\bar{k})$,
\begin{align}
    \rho=\sum_{m,n}\frac{c_{m,n}}{m!n!}(yk)^m(y\bar{k})^n \,,
\end{align}
where $m+n=2s-2$, which turns \eqref{rho} into
\begin{align}
    -4k^2\sum_{m,n}\frac{c_{m,n}}{m!n!}(m+1)(n+1)(yk)^m(y\bar{k})^n=-(k\partial)(\bar{k}\partial)\partial_C\phi\partial^C\phi \,.
\end{align}
We act on this with $(k\partial)^i(\bar{k}\partial)^j$ to get
\begin{align}
    c_{i,j}=\frac{(-1)^i}{(i+j+2)(i+j+1)}\frac{i!j!}{(2k)^{i+j+2}}{i+j+2\choose i+1}(k\partial)^{j+1}(\bar{k}\partial)^{i+1}\partial_C\phi\partial^C\phi\,.
\end{align}
Next, we introduce explicit indices and write $\rho$ as
\begin{align}
    \rho_{B(2s-2)}=\sum_{n=0}^{2s-1}\int \delta(z-z')R_{A(2s),B(2s-2)}\phi\fud{A(n)}{C}\phi^{A(2s-n)C} \,,
\end{align}
with
\begin{align}
    R_{A(2s),B(2s-2)}=\frac{1}{(2k)^{2s}}\sum_{i+j=2s-2}(-1)^i{2s\choose i+1}k_{A(j+1)}\bar{k}_{A(i+1)}k_{B(i)}\bar{k}_{B(j)} \,.
\end{align}
The $\phi$-component can be solved from the transverse part of the source $(\pl^{C} \phi \pl_{C} \phi)$ using the physical gauge propagator \eqref{PropPhysical}. This gives \begin{align}
    \begin{aligned}
    \phi_{B(2s-1)B'}(k,z')&=\sum_{n=1}^{2s-1}\int\langle \psi_{A(2s)}(-k,z)\phi_{B(2s-1)B'}(k,z')\rangle\phi\fud{A(n)}{C'}(k,z)\phi^{A(2s-n)C'}(k,z) \,.
    \end{aligned}
\end{align}
Now, we can assemble the ``electrostatic'' potential $\rho$ and the ``vector'' potential $\phi$ into $\Phi$. The complete gauge field then takes the form
\begin{align} \label{Phi}
    \Phi_{B(2s-1),B'}=\sum_{n=1}^{2s-1}\int\Big(\langle \psi_{A(2s)}\phi_{B(2s-1)B'}\rangle+\delta(z-z')R_{A(2s),B(2s-2)}\epsilon_{BB'}\Big)\phi\fud{A(n)}{C'}\phi^{A(2s-n)C'}\,.
\end{align}
The formula above is one step of the Berends--Giele recursion, which suffices to get to the four-point function. 

\subsubsection{Feynman/Lorenz gauge} 
By Feynman gauge in the first order theory we mean the result that we would obtain for $\langle \Psi \Phi\rangle$ in Feynman gauge in the second order theory, which, basically, corresponds to the kinetic term built from $\square$, i.e. without the $(\pl_\mu A^\mu)^2$-part. Let us set $\xi=0$. There is a cute trick \cite{Krasnov:2016emc,Krasnov:2021nsq} to absorb the Nakanishi-Lautrup field $B$ into $\Psi$, $\Psi^{A(2s-1),B}=\Psi^{A(2s-1)B}+\epsilon^{AB}B^{A(2s-2)}$, which allows to rewrite the kinetic term as
\begin{align}
    S_2&=\int \Psi^{A(2s-1),B} \nabla_{BC'}\Phi\fdu{A(2s-1),}{C'}\,. 
\end{align}
Here, the comma in $\Psi^{A(2s-1),B}$ means that there is no symmetry between $A(2s-1)$ and $B$, i.e. the tensor is reducible. The kinetic term is clearly invertible and the flat space propagator is
\begin{align}
    \langle \Psi^{A(2s-1),\bar{A}} \Phi^{B(2s-1),B'}\rangle_{\text{F}} &= \frac{1}{p^2} p^{\bar{A}B'} \epsilon^{AB}...\epsilon^{AB}\,.
\end{align}
One can think of this propagator as coming from the second order theory via the equations of motion $\slashed p \Phi=\Psi$, which is similar to the relation between $\langle \Psi \Phi\rangle $ and $\langle \Phi \Phi\rangle$ propagators in Chalmers--Siegel theory. The propagator $\langle \Phi \Phi\rangle$ is just $p^{-2}$ times $\epsilon$s, i.e. it is $p^{-2}\delta^{IJ}$, where $\delta^{IJ}$ is the unit operator on the space of $(2s-1,1)$-type spin-tensors. The Lorenz gauge propagator corresponds to integrating out $B$, which imposes $\nabla \cdot \Phi=0$. It can be obtained just by symmetrizing the indices on $\Psi$:
\begin{align}
    \langle \Psi^{A(2s)} \Phi^{B(2s-1),B'}\rangle_{\text{L}} &= \frac{1}{p^2} p^{A B'} \epsilon^{AB}...\epsilon^{AB}\,.
\end{align}
It is clearly $p$-transverse on the $\Phi$-leg.  
A Fourier transform gives the inhomogeneous part of the AdS${}_4$-propagator:
\begin{align}\notag
    \langle \Psi^{A(2s-1),\bar{A}}(-k,z) \Phi^{B(2s-1),B'}(k,z')\rangle_{\text{inh}}^{\text{F}} &=\nabla\fud{\bar{A}}{C'} \epsilon^{C'B'} \epsilon^{AB}...\epsilon^{AB}f(z-z')=\\ \label{PropFeyn}
    &= \epsilon^{AB(2s-1)} (k^{\bar{A}B'} +\epsilon^{\bar{A}B'}\pl_z ) f(z-z')\,.
\end{align}
As in flat space, the Lorenz gauge propagator is obtained via symmetrization
\begin{align}\label{PropLorenz}
    \langle \Psi^{A(2s)}(-k,z) \Phi^{B(2s-1),B'}(k,z')\rangle_{\text{inh}}^{\text{L}} 
    &= \epsilon^{AB(2s-1)} (k^{AB'} +\epsilon^{AB'}\pl_z ) f(z-z')\,.
\end{align}

\paragraph{Physical vs. Lorenz.} In order to manipulate the AdS/CFT correlators computed in the physical gauge, e.g. to see that the leading energy pole vanishes, it is advantageous to relate it to Lorenz gauge. In general, the difference between propagators in two gauges, say some gauge and the Feynman/Lorenz gauge, is a pure gauge term. 
However, in the physical gauge we have integrated out longitudinal modes by hand\footnote{In QED this is equivalent to integrating out $A_0$ via Gauss law, which generates an instantaneous interaction. This effectively removes the $G_{00}\sim 1/k^2$ component of the propagator and moves the result to the interaction part of the Lagrangian. } and we have to add the resulting instantaneous interaction back to the propagator
\begin{align} \label{gaugeVar}
\begin{aligned}
        \langle \psi_{A(2s)}\phi_{B(2s-1)B'}\rangle_\text{Phys}&=\langle \Psi_{A(2s)}(-k,z)\Phi_{B(2s-1),B'}(k,z')\rangle_\text{L}^{\text{inh}}+\\
        &\qquad+\nabla^{k,z'}_{BB'}\xi_{A(2s),B(2s-2)} +\eta_{A(2s),B(2s-1),B'}\delta(z-z')\,.
\end{aligned}
\end{align}
This relation can be solved for $\xi$ and $\eta$ and $\eta$ can be shown to compensate the instantaneous interactions (here we restrict ourselves to four-point functions for which only the quartic piece of the instantaneous interactions matters). We should stress that the relation is true with only the inhomogeneous part of the Lorenz gauge propagator. This means that whatever homogeneous part needs to be added to the Feynman/Lorenz gauge propagator it has to be a pure gauge term! This observation kills two birds with one stone. Firstly, we have a simple relation between physical and Lorenz gauge. Secondly, we have proved that all homogeneous terms in Lorenz gauge are pure gauge.\footnote{This is definitely not the case for the Lorenz-gauge propagator in Maxwell/Yang-Mills theory with Dirichlet/Neumann or generic mixed boundary conditions \cite{Skvortsov:2026gtq}. What makes it work is the self-duality. } 

We solve \eqref{gaugeVar} in the language of generating functions as it is more efficient. Note that we actually compute the difference between the Lorenz gauge propagator and the physical gauge one since the $B$-field is already gone by the time we impose the Coulomb gauge. We introduce $y^A$ and $z^A$, $w^{A'}$ to hide indices of $\Psi^{A(2s)}$ and $\Phi^{A(2s-1),A'}$, respectively. We need to solve
\begin{align}\notag
    -k \Pi_\text{o} f + \pl_z f \Pi_\text{e}&= (k^{AB'} f+ \epsilon^{AB'} \pl_z f) \epsilon^{AB(2s-1)}+ (k^{BB'}+\epsilon^{BB'}\pl_z) (\xi_1 f + \xi_2 \pl_z f)+\delta(z-z') \eta\,, 
\end{align}
where we displayed only the most important indices. We have also split the gauge parameter $\xi$ into two parts in accordance with the expected $z,z'$-dependence. Note that $\pl_z=-\pl_{z'}$ for the inhomogeneous terms. Some of the structures that appear when the indices are contracted with $y,z,w$ are
\besubeqs
\begin{align}
   \Pi_\text{e}&= \frac{1}{(4k^2)^s}[ (\brk y)^{2s} (kz)^{2s-1} (kw)+(k y)^{2s} (\brk z)^{2s-1} (\brk w)] \,, \\
   \Pi_\text{o}&= \frac{1}{(4k^2)^s}[ (\brk y)^{2s} (kz)^{2s-1} (kw)-(k y)^{2s} (\brk z)^{2s-1} (\brk w)]\,, \\
   (ykw)&= \tfrac12 [(ky)(\brk w)+(\brk y)(k w)]\,, \\
   (y \epsilon z)&= (zy)= \tfrac{1}{2k}[(ky) (\brk z)-(\brk y) (k z) ]\,, 
\end{align}
\esubeqs
where we used $k^{AB}=k^{(A}\brk^{B)}$ and $\epsilon^{AB}=\tfrac{1}{2k}(k^{A}\brk^B- k^B\brk^{A})$ in the last two lines. Isolating the terms in front of $f$, $\pl_z f$ and $\delta(z-z')$ we find
\besubeqs
\begin{align}
    -k \Pi_\text{o} &= (ykw) (y\epsilon z)^{2s-1} +(zkw) \xi_1 +(z\epsilon w) k^2\xi_2\,, \\
     \Pi_\text{e} &= (y\epsilon w) (y\epsilon z)^{2s-1} +(zkw) \xi_2 +(z\epsilon w) \xi_1\,, \\
     0&= \eta+2(z\epsilon w)\xi_2\,.
\end{align}
\esubeqs
The solution is
\besubeqs
\begin{align} \label{xisol}
    \xi_1&= \frac{1}{4^s k^{2s-1}}\frac{-\left(z_k y_{\bar{k}}\right){}^{2 s}+\left(y_k z_{\bar{k}}\right){}^{2 s}+\left(z_k y_{\bar{k}}+y_k z_{\bar{k}}\right) \left(z_k y_{\bar{k}}-y_k z_{\bar{k}}\right){}^{2 s-1}}{z_k z_{\bar{k}}}\,,\\
    \xi_2&=\frac{1}{4^s k^{2s}}\frac{ \left(\left(z_k y_{\bar{k}}\right){}^{2 s}+\left(y_k z_{\bar{k}}\right){}^{2 s}-\left(z_k y_{\bar{k}}-y_k z_{\bar{k}}\right){}^{2 s}\right)}{z_k z_{\bar{k}}}\,,\\
    \eta&= -2(z\epsilon w)\xi_2\,.
\end{align}
\esubeqs
Here, we abbreviated $y_k\equiv (ky)$, $y_{\bar{k}}=(\brk y)$ etc. It may not look obvious, but the generating functions are analytic and there are no poles. For $s=1$ we find $\xi_1=0$ and $\xi_2=\frac{y_k y_{\bar{k}}}{2 k^2}$, i.e. $\xi_2^{AA}\sim k^{AA}$. The index structure can be decoded as follows. One can recognize structures like $(ykz)$, $(y\epsilon z)$ in $\xi_{1,2}$. The numerator of $\xi_2$ is
\begin{align}
    \Pi_\text{e}^{A(2s),B(2s)} -\epsilon^{AB(2s)}&= k^{BB} \xi_2^{A(2s),B(2s-2)}= k^{BB}k^{AA} \chi_2^{A(2s-2),B(2s-2)}\,.
\end{align}
This structure is proportional to $k^{BB}=k^B\brk^{B}$ since it annihilates $k^{B(2s)}$ and $\brk^{B(2s)}$. The denominator cancels this $k^{BB}$-prefactor. In fact, since it is annihilated by $k^{A(2s)}$ and $\brk^{A(2s)}$ from the left as well, it is also proportional to $k^{AA}$, which can be seen by expanding the generating function. The l.h.s. above represents $(\mathrm{id} -\Pi_\text{e})$. Similarly, 
\begin{align}
    -k\Pi_\text{o}^{A(2s),B(2s)} -\epsilon^{AB(2s-1)}k^{AB}&=k^{BB} \xi_1^{A(2s),B(2s-2)}=k^{BB}k^{AA} \chi_1^{A(2s-2),B(2s-2)}\,.
\end{align}
Here, we can massage the l.h.s. either as $\slashed k (\mathrm{id} -\Pi_\text{e})$ or as $-(\mathrm{id} -\Pi_\text{e})\slashed k$. Note that $\slashed k$ preserves $k^A$ and $\brk^{A}$ since they are eigen vectors of $\slashed k$. 

\paragraph{Back to Feynman/Lorenz.} 
As we have just found out, we can only add a pure gauge homogeneous term to the inhomogeneous part of the propagator, which narrows down the ansatz\footnote{This was found to be true without the $B$-component. However, the BRST-invariance implies $0=s\langle \bar{c} \Phi_\mu\rangle=\langle B \Phi\rangle -\langle \bar{c} \pl_\mu c\rangle $, where we used $s=1$ as an example. In other words, $\langle B \Phi_\mu\rangle$ is a gradient. }
\begin{align}\notag
    \langle \Psi_{A(2s-1),\bar{A}} \Phi_{B(2s-1),B'}\rangle_{\text{hom}}^{\text{F}}=\nabla^{k,z'}_{BB'}e^{-k(z+z')}\xi_{A(2s-1),\bar{A},B(2s-2)}&= e^{-k(z+z')}k_{B}\brk_{B'} \xi_{A(2s-1),\bar{A},B(2s-2)}\,.
\end{align}
This pure gauge term must have no overlap with the physical directions, which is automatically true for the indices $B(2s-1),B'$ and have to be imposed on the $A(2s-1),\bar{A}$ indices. In terms of generating functions let us define the following decomposition
\besubeqs
\begin{align}
    \Phi&=(\bry \pl)\phi+ (\bry y)\rho\,,  & \Psi&=(\bry \pl)\psi+ (\bry y)B\,,\\
    \phi&= q+ (yky) \chi\,, & \psi&=\pi+ (yky)\sigma\,.
\end{align}
\esubeqs
Here, $\Phi$ and $\Psi$ are only linear in $\bry$ and $k\cdot q = k\cdot \pi=0$, i.e. $q$ and $\pi$ carry only the physical degrees of freedom. In particular, $q$ and $\pi$ are gauge invariant at the free level. The conditions we want to impose are $B(z=0)=0$ and $\chi(z'=0)=0$, which can be implemented as
\besubeqs
\begin{align}
    \epsilon_{A\bar{A}}\langle \Psi^{A(2s-1),\bar{A}}(-k,0) \Phi^{B(2s-1),B'}(k,z')\rangle&=0\,, \\ \langle \Psi^{A(2s-1),\bar{A}}(-k,z) \Phi^{B(2s-1),B}(k,0)\rangle k_{BB}&=0\,.
\end{align}
\esubeqs
These are the usual conditions for the Dirichlet problem, see Appendix \ref{app:Unphysical}. A supporting argument is that these boundary conditions together with the homogeneous term being pure gauge fix the propagator. With a bit of luck one finds
\begin{align} 
\begin{aligned}\label{xiprime}
       \xi_{A(2s-1),\bar{A},B(2s-2)}&= \xi'_{A(2s-1),B(2s-2)} \brk_{\bar{A}}\,,\\
    \xi'&= \frac{1}{(2k)^{2s-1}} \frac{\left(z_k y_{\bar{k}}-y_k z_{\bar{k}}\right){}^{2 s-1}-\left(z_k y_{\bar{k}}\right){}^{2 s-1}}{z_{\bar{k}}}\,. 
\end{aligned}
\end{align}
Note that at tree level the ghosts can be dropped and, hence, the $\bar{A}$-index is always symmetrized with the rest of $A(2s-1)$ since they are always contracted with the vertices that are symmetric. The tensor structure here is
\begin{align}
    -\epsilon^{AB(2s-1)} - \frac{1}{(2k)^{2s-1}}\brk^{A(2s-1)} k^{B(2s-1)}= \brk^B \xi^{A(2s-1),B(2s-2)}=k^A \brk^B \chi^{A(2s-2),B(2s-2)}\,.
\end{align}
It is the structure that is annihilated by $\brk_{B(2s-1)}$ from the right and by $k_{A(2s-1)}$ from the left, hence, it has to be proportional to $\brk^B k^A$. One can introduce two projectors:
\begin{align}
    \Pi_{\brk k}^n&=\frac{1}{(4k^2)^{n}} \brk^{A(n)} k^{B(n)}\,, & \Pi_{k\brk}^n&=\frac{1}{(4k^2)^{n}} k^{A(n)} \brk^{B(n)}\,.
\end{align}
The even and odd projectors are made of these
\begin{align}
    \Pi_\text{e}&= \Pi_{\brk k}+\Pi_{k\brk}\,, & \Pi_\text{o}&= \Pi_{\brk k}-\Pi_{k\brk}\,.
\end{align}
Being careful about the position of indices $-(\mathrm{id}+\Pi_{\brk k})$ subtracts one physical direction from the total space. Note that the pure gauge piece does not have any overlap with the physical states as it is multiplied by $k^B\brk^{B'}$. 

Let us make one observation that will be important for the Feynman and physical gauges to give the same results. There is a purely algebraic identity
\begin{align}
    \xi_1^{A(2s),B(2s-2)}-k\xi_2^{A(2s),B(2s-2)}=\xi^{A(2s-1),A,B(2s-2)} \,,
\end{align}
which is obvious in terms of generating functions. When one computes the four-point function in the physical gauge with the help of the relation \eqref{gaugeVar}, the pure gauge part can be integrated by parts. Since the external legs are on-shell, only the boundary term survives:
\begin{align}
    \epsilon^{BB'}[\xi_1 f(z-z') +\xi_2 \pl_z f(z-z')]\Big|_{z'=0}&=f(z)[\xi_1-k\xi_2] \epsilon^{BB'}\,.
\end{align}
Here, we omit the indices on $\xi$s, but $\epsilon^{BB'}$ comes from $\nabla^{BB'}\ni -\epsilon^{BB'} \pl_{z'}$ and the sign is due to $z'=0$ being the lower limit. Note that $f(z)=-e^{-kz}/(2k)$. Therefore, we see that, when the inhomogeneous part of the Feynman propagator is used as a references point, the pure gauge contribution of the physical gauge and of the full Feynman propagator agree with each other on the boundary.

\subsection{Boundary-to-bulk propagators, two-point functions} 
Boundary-to-bulk propagators are just limits of the bulk-to-bulk ones. They have already been implicitly found via the Fefferman-Graham analysis as the regular solutions. In the axial or physical gauges we have
\besubeqs
\begin{align}
    \Phi^{A(2s-1),A'}(k,z)&=\alpha_s^+ \epsilon_+^{A}...\epsilon_+^{A}\epsilon_+^{A'}e^{-kz}\sim  \brk^{A}...\brk^{A}\brk^{A'} e^{-kz}\,,\\
    \Psi^{A(2s)}(k,z)&= \alpha_s^- \epsilon_-^{A}....\epsilon_-^{A}e^{-kz} \sim k^{A}...k^{A} e^{-kz}\,.
\end{align}
\esubeqs
In the Feynman gauge the equations are weaker and one can, in principle, use 
\besubeqs
\begin{align}
    \Phi^{A(2s-1),A'}(k,z)&=\alpha_s^+  (qk)^{-s}q^A...q^A\brk^{A'}e^{-kz}\,,\\
    \Psi^{A(2s-1),\bar{A}}(k,z)&= \alpha_s^- (\brk\bar{q})^{-s}\bar{q}^A...\bar{q}^A  k^{\bar A}e^{-kz} \,,
\end{align}
\esubeqs
where $q^A$, $\bar{q}^{A}$ are reference spinors. With $q^A=\brk^A$, $\bar{q}^A=k^A$ we get back the propagators in the physical/axial gauge. The latter have fixed helicity. For generic $q$, $\bar{q}$ the Feynman gauge propagators do not correspond to CFT states of definite helicity and we shall not use them.

\paragraph{Two-point functions.} Two-point functions are somewhat subtle in AdS/CFT and require appropriate boundary terms being added to the action. These are the boundary terms to make the action stationary plus, possibly, some additional boundary terms that lead to contact contributions to correlators. The free action for chiral fields is first order and vanishes on-shell (similarly to the Dirac action). It is already stationary (as different from the Dirac action \cite{Arutyunov:1998ve, Henneaux:1998ch,Papadimitriou:2004ap}) and does not require any boundary terms. However, one can add boundary terms of type $\Psi \Phi$. Evaluating such terms first at $z=\epsilon>0$ and taking the limit $\epsilon\rightarrow0$, the only combination that survives is $\Psi_-(z) \Phi_+(z)$. If we forget about the helicity discrimination present in the first order formulation, the corresponding contact term in a two-point function is of type
\begin{align}
    \langle \phi_{A(2s)} J_{B(2s)}\rangle \sim \delta^3(x-y) \,.
\end{align}
This is a generalization of the statement that the two-point function of an operator and its ``shadow'' is a contact term, $\langle O_\Delta O_{d-\Delta}\rangle \sim \delta^3 (x-y)$. On the other hand, the bulk-to-bulk propagators \eqref{propPhysicalComp} contain $\sign(z-z')$, which makes the limit $z,z'\rightarrow0$ of the propagator depend on the order of limits. The difference is a contact term, which is a legitimate ambiguity present in the formalism.

On the other hand, it was shown in \cite{Chowdhury:2024dcy} that different choices of reference spinors lead to two-point functions with different helicity structures. One subtlety is that the concept of helicity is tied to momentum space and, hence, may obscure locality in space-time. This is clearly visible in the example of two-point functions of a conserved current. There are two structures: nonlocal parity-even, $a k(\delta_{ij}-k_ik_j/k^2)$ and local parity-odd, $b\epsilon_{ijn}k^n$. In the helicity basis they contribute on equal footing $k(a+b)$, $k(a-b)$ to $\langle ++\rangle$, $\langle -- \rangle$. At present, more input is needed to fix two-point functions since they are given solely by contact terms.

\subsection{Three-point functions}
\label{sec:}
We will compute the correlation functions using generating functions. The generating functions for boundary-to-bulk propagators read
\begin{align}
    \Phi_s(k_i,z)&=\alpha_s(k_i)(\bar{i}^Ay_A)^{2s_i-1}(\bar{i}^Bw_B)e^{-k_iz}\,, & \Psi_s(k_i,z) &=\alpha_s(k_i)(i^Ay_A)^{2s_i}e^{-k_iz}\,,
\end{align}
with the canonical normalization $\alpha_s(k_i)=\frac{1}{(2k_i)^{s_i}}$. However, the most natural normalization is the one that leads to the simplest generating function of the propagators:
\begin{align}
    \phi&= \sum_s \phi_s(k_i,z)=\alpha_+\exp[(\bar{i}^Ay_A)\alpha_i  -k_iz]\,, & \Psi=\alpha_-\exp[(i^Ay_A)\alpha_i-k_iz]\,,
\end{align}
where $\alpha_i=\alpha(k_i)=1/\sqrt{2k_i}$ and $\Phi=(\bry \pl)\phi$ is the symmetric component of $\Phi$. Apart from normalizing the propagators, $\alpha_i$ can be useful as a bookkeeping device for spin: the order $\alpha_i^{2s_i}$ terms in the Taylor expansion of a correlator corresponds to a spin-$s_i$ operator on the CFT side. 

To derive an expression for the vertex, let us take the field strength \eqref{nfactor} (with $N$-factor removed)
\begin{align}
    (yFy)&= (y\nabla \plb) \Phi+(\plb_{C'} \Phi) (\plb^{C'} \Phi)\,.
\end{align}
The corresponding cubic vertex, coming from $\langle \Psi| (yFy)\rangle$, can be written as a poly-differential operator (a coupling constant $2g$ is introduced)
\begin{align}
    V_3(\partial_1,\bar{\partial}_1;\partial_2,\bar{\partial}_2;\partial)&=2ge^{(\pl_1+\pl_2)\pl_3} \plb_{C'}^1 \Phi(y_1,\bry_1) \plb^{C'}_2 \Phi(y_2,\bry_2)\Psi(y_3)\Big|_{y_i=0,\bry_i=0}\,.
\end{align}
If $\Phi$ is an external line, it is symmetric $\Phi=(\bry \pl)\phi$ and one gets an even simpler expression 
\begin{align}
    V_3(\partial_1;\partial_2;\partial)&=2ge^{(\pl_1+\pl_2)\pl_3}\pl_C^1 \pl_2^C \phi(y_1) \phi(y_2)\Psi(y_3)\Big|_{y_i=0}\,.
\end{align}
In $\text{AdS}_4$, four-momentum is not conserved. Three-momentum is conserved in the boundary directions, while in the radial direction the non-conservation is expressed by the energy $E$, i.e.
\begin{align} \label{E}
    1^A\bar{1}^{A'}+2^A\bar{2}^{A'}+3^A\bar{3}^{A'}&=E\epsilon^{AA'} \,, & k_1+k_2+k_3&=E\,.
\end{align}
The flat limit is obtained at $E\rightarrow 0$, where four-momentum conservation is restored. The (generating function of) three-point correlator is
\begin{align}
    \mathcal{W}_3&=\int V(\partial_1;\partial_2;\partial)\Phi(k_1,z)\Phi(k_2,z)\Psi(k_3,z)=-\frac{2g\alpha_+^2\alpha_-\alpha_1\alpha_2}{E}\langle \bar{1}\bar{2} \rangle e^{\alpha_1\alpha_3\langle\bar{1}3\rangle}e^{\alpha_2\alpha_3\langle\bar{2}3\rangle}\,.
\end{align}
It is useful to consider the correlation function for a specific spin-configuration, where we should remind ourselves that we are interested in spin-configurations that contain only nonzero integer spins, which implies that we require an overall factor $\Pi_{i=1}^3\alpha_i^{2s_i}$. We also introduce the spin-specific normalization $\mathcal{N}_{s_1,s_2,s_3}=\frac{\alpha_+^2\alpha_-\alpha_1^{2s_1}\alpha_2^{2s_2}\alpha_3^{2s_3}}{(2s_1-1)!(2s_2-1)!}$. The correlator reads\footnote{The angle bracket is defined as $\langle ij \rangle= i^Aj_A$ and is used for both barred and unbarred spinors, potentially even mixing them.}
\begin{align}
    \begin{aligned}
    \mathcal{W}_3^{s_1s_2s_3}&=-\frac{2g\mathcal{N}_{s_1,s_2,s_3}}{E}\langle \bar{1}\bar{2} \rangle \langle \bar{1}3\rangle^{2s_1-1}\langle \bar{2}3\rangle^{2s_2-1}\delta_{s_1+s_2-s_3,1}=\\
    &=\mathcal{N}_{s_1s_2s_3}\langle \bar{1}3 \rangle^{2s_1-2}\langle \bar{2}3 \rangle^{2s_2-2}\delta_{s_1+s_2-s_3,1}\mathcal{W}_3^\text{SDYM}\,,
    \end{aligned}
\end{align}
where $\mathcal{W}_3^\text{SDYM}$ is the SDYM three-point correlator with its normalization stripped off: 
\begin{align}\label{SDYM3ptAxial}
    \mathcal{W}_3^{\text{SDYM}} = -\frac{2g}{E}\langle \bar{1}3 \rangle \langle \bar{2}3 \rangle\langle \bar{1}\bar{2} \rangle\,.
\end{align}
Hence, the HS-SDYM three-point function is just the SDYM one dressed with a higher-spin factor. The definition for the energy \eqref{E} can be used to derive the relations
\begin{align} \label{momConsCorrections}
    \langle \bar{1}3 \rangle &= \frac{2k_1 \langle \bar{1}\bar{2} \rangle-E\langle \bar{1}\bar{2} \rangle}{\langle \bar{2}\bar{3} \rangle} \,, & \langle \bar{2}3 \rangle &= -\frac{2k_2 \langle \bar{1}\bar{2} \rangle-E\langle \bar{1}\bar{2} \rangle}{\langle \bar{1}\bar{3} \rangle}\,.
\end{align}
With their help we bring the correlator to a more familiar form
\begin{align}
    \mathcal{W}_3^{s_1s_2s_3}=\frac{2g^2\mathcal{N}_{s_1s_2s_3}}{E}(2k_1)^{2s_1-1}(2k_2)^{2s_2-1}\frac{\langle \bar{1}\bar{2}\rangle^{2s_1+2s_2-1}}{\langle \bar{1}3\rangle^{2s_2-1}\langle \bar{2}3\rangle^{2s_1-1}}\delta_{s_1+s_2-s_3,1} + \mathcal{O}(E^0)\,.
\end{align}
We isolate the energy pole by taking the flat limit
\begin{align}
    \lim_{E\rightarrow 0} E\mathcal{W}_3 &\sim \mathcal{A}_3\,,
\end{align}
with $\mathcal{A}_3$ the flat space amplitude. The proportionality constant is simply the normalizations of the boundary-to-bulk propagators.

\subsection{Four-point functions}
\label{sec:}
We will compute the four-point $s$-channel correlator in the physical gauge. The correlator is given by
\begin{align}\label{s-channel}
    \begin{aligned}
        \mathcal{W}_4^s &= \int \Phi(k_1,z)\Phi(k_2,z)V(\partial_1;\partial_2;\partial_v)\langle \Psi(y_v)\Phi(y_w,\bar{y}_w)\rangle V(\partial_w,\bar{\partial}_w;\partial_3,\partial_4)\Phi(k_3,z')\Psi(k_4,z') =\\
        &= 4g^2\alpha_+^3\alpha_-\alpha_1\alpha_2\langle \bar{1}\bar{2} \rangle e^{\alpha_3\alpha_4\langle\bar{3}4\rangle}\int e^{-(k_1+k_2)z}\langle \Psi(-\alpha_1\bar{1}-\alpha_2\bar{2})\Phi(\alpha_44,\alpha_3\bar{3})\rangle e^{-(k_3+k_4)z'}\,.
    \end{aligned}
\end{align}
Here, we should remember the fact that $\langle \Psi(y_v)\Phi(y_w,\bar{y}_w)\rangle $ is linear in $\bar{y}_w$.
The bulk-to-bulk propagator can be decomposed as\footnote{Here we ignore the $\eta\delta(z-z')$ term, as it is canceled by the instantaneous interaction.}
\begin{align}
    \langle \Psi\Phi \rangle = \langle \Psi\Phi \rangle_{\text{L}}^{\text{inh}} + \nabla\xi \,,
\end{align}
where $\xi$ can be found in \eqref{xisol} and\footnote{Note that we included $s=0$ and half-integer spins in the summation to improve the simplicity of the generating function. Eventually, one should remember to project onto positive integer spins. This should also take care of the pole that may be present in some generating functions. One can also keep the half-integer spins as HS-SDYM has a straightforward super-symmetric extension.}
\begin{align}
    \begin{aligned}
        \langle \Psi(y_v)\Phi(y_w,\bar{y}_w) \rangle^\text{inh}_\text{L} &= \sum_{s\in\mathbb{N}_0/2}\frac{(y_v\epsilon y_w)^{2s-1}}{(2s)!}\Big((y_vk\bar{y}_w)f+(y_v\epsilon \bar{y}_w)\partial_z f\Big) =\\
        &= \frac{e^{(y_v\epsilon y_w)}}{(y_v\epsilon y_w)}\Big((y_v\epsilon \bar{y}_w) f + (y_vk\bar{y}_w)\partial_z f\Big) \,.
    \end{aligned}
\end{align}
Accordingly, we decompose $\mathcal{W}_4^s$ into
\begin{align} \label{decomp}
    \mathcal{W}_4^{s}=\mathcal{W}_\text{L}^{\text{inh}}+\mathcal{W}^{\nabla\xi} \,.
\end{align}
Like for the three-point functions, we express the non-conservation of four-momentum as the energy $E$,
\begin{align}
    k_1+k_2+k_3+k_4=E\,.
\end{align}
Moreover, we denote the non-conservation of four-momentum in the left and right interaction vertex as
\begin{align}
    E_\text{L}&= k_1+k_2+k \,, & E_\text{R}&=k_3+k_4+k\,,
\end{align}
respectively. The three-momentum of the internal line is $k^{AA'}=k_1^{AA'}+k_2^{AA'}$. This allows us to write the integrals relevant to the four-point functions as
\begin{align}
    \begin{aligned}
        \int e^{-(k_1+k_2)z}fe^{-(k_3+k_4)z'} &= -\frac{E+2k}{2kEE_\text{L}E_\text{R}} \,,\\
        \int e^{-(k_1+k_2)z}\partial_zfe^{-(k_3+k_4)z'} &= \frac{E_\text{R}-E_\text{L}}{2EE_\text{L}E_\text{R}} \,,\\
        \int e^{-(k_1+k_2)z}f\Big|_{z'=0} &= -\frac{1}{2kE_\text{L}}\,.
    \end{aligned}
\end{align}
For $\langle\Psi\Phi\rangle_\text{L}^\text{inh}$, the integral in the last line of \eqref{s-channel} yields
\begin{align}
    \int e^{-(k_1+k_2)z}\langle \Psi(y_v)\Phi(y_w,\bar{y}_w)\rangle_\text{L}^\text{inh} e^{-(k_3+k_4)z'} &= -2k\frac{e^{(y_v\epsilon y_w)}}{(y_v\epsilon y_w)}\Big((y_vp_{34}\bar{y}_w)-\frac{E}{2k}(y_vp\bar{y}_w)\Big)\,,
\end{align}
where $p_{34}^{AA'}=3^A\bar{3}^{A'}+4^A\bar{4}^{A'}$ and $p^{AA'}=k^{AA'}+k\epsilon^{AA'}=k^A\bar{k}^{A'}$. It then follows that
\begin{align} \label{WFinh}
    \begin{aligned}
        \mathcal{W}_\text{L}^{\text{inh}}&=-\frac{4g^2\alpha_+^3\alpha_-\alpha_1\alpha_2\alpha_3}{EE_\text{L}E_\text{R}} \langle \bar{1}\bar{2} \rangle \Big(\alpha_1\langle\bar{1}4\rangle\langle \bar{3}4\rangle+\alpha_2\langle \bar{2}4 \rangle\langle \bar{3}4 \rangle - \frac{E}{2k}\big(\alpha_1\langle \bar{1}|k\bar{k}|\bar{3}\rangle + \alpha\langle \bar{2}|k\bar{k}|\bar{3}\rangle\big)\Big)\times\\
        &\times \frac{e^{(\alpha_1\alpha_4\langle \bar{1}4 \rangle +\alpha_2\alpha_4\langle \bar{2}4\rangle})}{\alpha_1\alpha_4\langle \bar{1}4\rangle + \alpha_2 \alpha_4 \langle \bar{2}4\rangle}e^{\alpha_3\alpha_4\langle\bar{3}4\rangle}\,.
    \end{aligned}
\end{align}
Let us again evaluate the correlation function for a certain spin-configuration. We need to be careful to remove spin-configurations with half-integer spins. This means that we must get an overall factor $\Pi_{i=1}^4\alpha_i^{2s_i}$, which requires different orders in the expansion of the generating functions to survive for the terms in the parentheses. The spin-specific correlation function reads
\begin{align}
    \begin{aligned}
        \mathcal{W}^{s,\text{L}}_{s_1s_2s_3,s_4} &=-\frac{4g^2\mathcal{N}_{s_1s_2s_3s_4}}{EE_\text{L}E_\text{R}}\langle \bar{1}\bar{2} \rangle\langle \bar{1}4 \rangle^{2s_1-2} \langle \bar{2}4\rangle^{2s_2-2}\langle \bar{3}4 \rangle^{2s_3-1}\Big(\langle \bar{1}4 \rangle \langle \bar{2}4\rangle\langle \bar{3}4 \rangle+\\
        &-\frac{E}{2k}\frac{(2s_1-1)\langle \bar{1}|k\bar{k}|\bar{3}\rangle+(2s_2-1)\langle \bar{2}|k\bar{k}|\bar{3}\rangle}{2s_1+2s_2-2}\Big) \delta_{s_1+s_2+s_3-s_4,2} \,,
    \end{aligned}
\end{align}
with the normalization $\mathcal{N}_{s_1s_2s_3s_4}=\frac{\alpha_+^3\alpha_-\alpha_1^{2s_1}\alpha_2^{2s_2}\alpha_3^{2s_3}\alpha_4^{2s_4}}{(2s_1-1)!(2s_2-1)!(2s_3-1)!}$. The first term is the energy pole of the four-point $s$-channel correlator and it is just the SDYM one multiplied by a higher-spin factor. The $t$-channel diagram is easily obtained by swapping $1\leftrightarrow 3$, $\bar{1} \leftrightarrow \bar{3}$ and $k_1\leftrightarrow k_3$.\footnote{$E_\text{L}$, $E_\text{R}$, $k$ and $\bar{k}$ contain objects whose labels should be swapped. It is useful to label them according to which channel they belong to, e.g. $E_\text{L}^s$, $k_s$.} Due to the overall factor compared to SDYM, it is easy to see that the energy pole vanishes for the same reasons as for SDYM when the $s$-channel and $t$-channel are added together.

To evaluate $\mathcal{W}^{\nabla\xi}$, we write the generating function for $(\nabla\xi)(y_v;y_w,\bar{y}_w)$ as
\begin{align}
    \begin{aligned}
        (\nabla\xi)(y_v;y_w,\bar{y}_w)&=\sum_{s\in\mathbb{N}_0/2}\frac{1}{(2s)!}\nabla_{BB'}\xi_{A(s),B(s-2)}y_v^A\dots y_v^A y_w^B\dots y_w^B\bar{y}_w^{B'}=\\
        &=(y_w\nabla \bar{y}_w)\sum_{s\in\mathbb{N}_0/2}\frac{1}{(2s)!}\xi_{A(s),B(s-2)}y_v^A\dots y_v^A y_w^B\dots y_w^B \,,
    \end{aligned}
\end{align}
which replaces the bulk-to-bulk propagator in \eqref{s-channel}. The $z'$-integral can be integrated by parts.\footnote{The covariant derivative belongs to the $\Phi$-leg of the bulk-to-bulk propagator, so $\nabla\xi=\nabla_{k,z'}\xi$.} When the covariant derivative hits the exponential that originates from the boundary-to-bulk propagators, we find the equations of motion for the external legs:
\begin{align}
    (4\nabla\bar{3})e^{-(k_3+k_4)z'}=(3_A\bar{3}_{A'}+4_A\bar4_{A'})4^A\bar{3}^{A'}e^{-(k_3+k_4)z'}=0\,.
\end{align}
Thus, only the total $z'$-derivative survives after imposing three-momentum conservation and $\nabla\xi$ must be evaluated at $z'=0$, which is given by \eqref{xiprime}. Its generating function reads
\begin{align}
    \begin{aligned}
        &(\nabla\xi)(y_v;y_w,\bar{y}_w)\Big|_{z'=0} = -\frac{e^{-kz}}{2k}\big((\nabla\xi_1)-k(\nabla\xi_2)\big)(y_v;y_w,\bar{y}_w)=\\
        &=(y_w\nabla\bar{y}_w)\frac{(\bar{k}y_v)}{(\bar{k}y_w)}\Big(\frac{e^{\alpha^2(\bar{k}y_v)(ky_w)}}{(\bar{k}y_v)(ky_w)}-\frac{e^{\alpha^2\big((\bar{k}y_v)(ky_w)-(ky_v)(\bar{k}y_w)\big)}}{(\bar{k}y_v)(ky_w)-(ky_v)(\bar{k}y_w)}\Big)e^{-kz}\,,
    \end{aligned}
\end{align}
with $\alpha=1/\sqrt{2k}$ and we used $k^{AA'}-k\epsilon^{AA'}=\bar{k}^Ak^{A'}$. As explained above, due to integration by parts and three-momentum conservation, only the $\epsilon$-component of the covariant derivative needs to be considered, i.e. we need to only evaluate $\big((\epsilon\xi_1)-(\epsilon\xi_2)\big)(y_v;y_w,\bar{y}_w)$. Finally, $\mathcal{W}^{\nabla\xi}$ is given by
\begin{align}
    \begin{aligned}
        \mathcal{W}^{\nabla\xi} &= -\frac{4g^2\alpha_+^3\alpha_-\alpha_1\alpha_2}{2kE_\text{L}}\langle \bar{1}\bar{2} \rangle e^{\alpha_3\alpha_4\langle \bar{3}4 \rangle}((\epsilon\xi_1)-k(\epsilon\xi_2))(-\alpha_1\bar{1}-\alpha_2\bar{2};\alpha_44,\alpha_3\bar{3}) =&\\
        &=-\frac{4g^2\alpha_+^3\alpha_-\alpha_1\alpha_2\alpha_3}{E_\text{L}}\langle \bar{1}\bar{2} \rangle\langle \bar{3}4\rangle\frac{\alpha_1\langle \bar{1}\bar{k}\rangle+\alpha_2\langle \bar{2}\bar{k}\rangle}{\langle 4\bar{k}\rangle}e^{\alpha_3\alpha_4\langle\bar{3}4\rangle}\Big(\frac{e^{-\alpha^2\alpha_4(\alpha_1\langle \bar{1}4\rangle+\alpha_2\langle \bar{2}4\rangle)\langle 4k \rangle}}{\alpha_4(\alpha_1\langle \bar{1}4\rangle+\alpha_2\langle \bar{2}4\rangle)\langle 4k \rangle}+\\
        &-\frac{e^{-\alpha^2\alpha_4\big(\alpha_1\langle \bar{1}\bar{k}\rangle+\alpha_2\langle \bar{2}\bar{k}\rangle)\langle 4k\rangle
        -(\alpha_1\langle \bar{1}k\rangle +\alpha_2\langle \bar{2}k\rangle)\langle 4\bar{k} \rangle\big)}}{\alpha_4\big(\alpha_1\langle \bar{1}\bar{k}\rangle+\alpha_2\langle \bar{2}\bar{k}\rangle)\langle 4k\rangle
        -(\alpha_1\langle \bar{1}k\rangle +\alpha_2\langle \bar{2}k\rangle)\langle 4\bar{k}\rangle\big)}\Big)\,.
    \end{aligned}
\end{align}
For a specific helicity configuration we get
\begin{align}
    \begin{aligned}
        &\mathcal{W}^{\nabla\xi}_{s_1s_2s_3s_4} = \frac{4g^2\mathcal{N}_{s_1s_2s_3s_4}}{E_\text{L}}\frac{\alpha^{4s_1+4s_2-2}}{2s_1+2s_2-2}\langle \bar{1}\bar{3} \rangle \langle \bar{3}4\rangle^{2s_3}\delta_{s_1+s_2+s_3-s_4,2}\times\\
        &\times\Big(\sum_{\substack{n=0,\dots,2s_1-2\\m=0,\dots,2s_2-1\\(n,m)\neq(0,0)}}\tfrac{(2s_1-1)!(2s_2-1)!}{n!m!(2s_1-n-2)!(2s_2-m-1)!}\langle\bar{1}\bar{k}\rangle^{2s_1-n-1}\langle\bar{2}\bar{k}\rangle^{2s_2-m-1}\langle \bar{1}k\rangle^n\langle\bar{2}k\rangle^{m}\langle 4\bar{k}\rangle^{n+m-1}\langle 4k \rangle^{2s_1+2s_2-n-m-3}+\\
        &+\sum_{\substack{n=0,\dots,2s_1-1\\m=0,\dots,2s_2-2\\(n,m)\neq(0,0)}}\tfrac{(2s_1-1)!(2s_2-1)!}{n!m!(2s_1-n-1)!(2s_2-m-2)!}\langle\bar{1}\bar{k}\rangle^{2s_1-n-1}\langle\bar{2}\bar{k}\rangle^{2s_2-m-1}\langle \bar{1}k\rangle^n\langle\bar{2}k\rangle^{m}\langle 4\bar{k}\rangle^{n+m-1}\langle 4k \rangle^{2s_1+2s_2-n-m-3}\Big)
    \end{aligned}\notag
\end{align}
Using that
\begin{align}
    E_\text{L}E_\text{R}=-\langle 12 \rangle \langle \bar{1}\bar{2} \rangle +EE_\text{L} \,,
\end{align}
we observe that in the flat limit
\begin{align}
    \lim_{E\rightarrow 0} E\mathcal{W}_4^s=\lim_{E\rightarrow 0} E\mathcal{W}_\text{L}^\text{inh}=4g^2\mathcal{N}_{s_1s_2s_3s_4}\langle \bar{1}4 \rangle^{2s_1-2} \langle \bar{2}4 \rangle^{2s_2-2} \langle \bar{3}4 \rangle^{2s_3-2}\frac{\langle \bar{3}\bar{4} \rangle}{\langle\bar{1}\bar{2}\rangle} \langle \bar{1}4 \rangle \langle \bar{2}4\rangle \langle \bar{3}4 \rangle \sim\mathcal{A}_4^s\,,
\end{align}
where $\mathcal{A}_4^s$ is the flat space $s$-channel amplitude. After adding up the $s$- and $t$-channels the flat limit of the four-point amplitude vanishes for generic kinematics, see e.g. \cite{Ponomarev:2017nrr,Guarini:2026vds}, and does not vanish for the collinear one \cite{Serrani:2026azw}. Let us also note that the Dirichlet four-point amplitude in Yang-Mills theory was computed in the spinor-helicity language in \cite{Armstrong:2020woi}, see also \cite{Chowdhury:2024dcy,Skvortsov:2026gtq} where the relation to SDYM is discussed.  

\section{Discussion and Conclusions}
\label{sec:conclusions}
In the paper we proposed the general principles of self-dual holography, which are based on the Fefferman-Graham expansion of the free equations for massless (self-dual) fields in the chiral description. There are two fields: $\Phi$ and $\Psi$ and one of them is dual to a gauge field $h_-$ on the boundary, the other is dual to a current $j_+$. However, each of them is dual to only a positive-helicity or negative-helicity ``half'' of the corresponding CFT field. By constructing the conserved current from the gauge field via its ``Cotton'' tensor $C(h_-)$ we can assemble a complete spin-$s$ current $j=\{j_+,j_-\}$ on the boundary:
\begin{align}
    &\Phi &&\longleftrightarrow&& j_+ \oplus j_-=C(h_-) &&  \longleftarrow &&h_- &&\longleftrightarrow&& \Psi\,.
\end{align}

Some obvious extensions of the present work include the computation of all $3$- and $4$-point functions in Chiral HiSGRA. Note that the correlators/amplitudes of SDYM/SDGR are contained in those of Chiral theory. To study $4$-point and higher order correlators it would be important to rethink the Berends--Giele approach in AdS. The relative simplicity of the bulk-to-bulk propagators should allow one to compute one-loop amplitudes. At one-loop, one should first verify that there are no UV-divergences in Chiral theory in (A)dS space, see \cite{Skvortsov:2020gpn,Skvortsov:2020wtf} for the corresponding result in flat space. One-loop corrections should give anomalous dimensions to CFT operators. It would be important to complete the definition of self-dual CFTs  \cite{Aharony:2024nqs,Jain:2024bza} and compute at least $3$- and $4$-point correlators to check the self-dual AdS/CFT duality. As a step towards this goal, it can be interesting to analyze the just obtained three- and four-point correlators from the CFT vantage point, e.g. to work out the conformal block expansion. 

All self-dual theories in flat space can be uniformized in the light-cone gauge as SDYM with the color algebra replaced by an appropriate kinematical algebra \cite{Monteiro:2011pc,Ponomarev:2017nrr,Monteiro:2022lwm,Serrani:2025oaw}. Via DDM ordering this reduces all tree-level (and one-loop) diagrams to those of SDYM \cite{Serrani:2026azw}. However, the idea of the kinematic algebra has not yet been generalized to self-dual theories on (A)dS, see, however, an example of SDGR \cite{Lipstein:2023pih} and the developments of the (A)dS light-cone approach in \cite{Metsaev:2018xip,Neiman:2023bkq, Neiman:2024vit, Chowdhury:2024dcy,Kozaki:2025jrj}. Therefore, new tools are needed to compute (A)dS/CFT correlators efficiently.

In flat space and in Feynman gauge the propagator factorizes into the SDYM propagator and several factors of $\epsilon^{AB}$, which considerably simplifies the calculations. For gauge interactions of HS-SDYM, the $n$-point amplitudes are those of SDYM times a higher-spin dressing factor \cite{Serrani:2026azw}. As we have shown in the paper, this is no longer true for AdS/CFT correlators as they do depend on spin in a nontrivial way, which is due to a homogeneous term in the propagator that is responsible for the appropriate boundary conditions. Nevertheless, the Lorenz/Feynman and physical gauges do lead to the same result, as is manifest in our calculations.

Since the chiral description originates in twistor theory, one can hope that the self-dual holography can be formulated directly in twistor space, thereby bypassing the usual spacetime picture. Some self-dual higher-spin theories have already been formulated on twistor space \cite{Tran:2021ukl,Adamo:2022lah,Herfray:2022prf,Tran:2022tft,Mason:2025pbz}. A closely related structure is the cosmological Grassmannian \cite{Arundine:2026fbr}. 

At the level of cubic interactions chiral and anti-chiral interactions provide a complete basis of interactions. Therefore, it is possible to compute all three-point functions in Chern--Simons matter theories from Chiral HiSGRA, as was done in the light-cone gauge in \cite{Skvortsov:2018uru}, but it is difficult to directly compare these results to covariant ones. It was also shown in \cite{Skvortsov:2018uru} that the $3d$-bosonization at the level of three-point functions is closely related to the existence of Chiral HiSGRA, see \cite{Skvortsov:2022wzo} for the four-point argument.

Going beyond Chiral theory one must confront the genuine nonlocality of the AdS dual of vector models. Nevertheless, an expansion over the Chiral theory, which is similar to the MHV-expansion, can help to quantify the nonlocality. For example, the scalar four-point function vanishes in (anti)-Chiral theories and, hence, its nonlocality comes directly from the chiral--antichiral exchange, see \cite{De:2026shn} for the recent discussion of this nonlocality.

\section*{Acknowledgment}
We are indebted to Simone Giombi for many very useful discussions during the earlier stage of this project. The work of E. S. and R. van D. was partially supported by the European Research Council (ERC) under the European Union’s Horizon 2020 research and innovation programme (grant agreement No 101002551). E.S. is a research associate of the Fonds de la Recherche Scientifique – FNRS. 
\appendix 

\section{Boundary conditions in gauged-fixed Maxwell theory}
\label{app:BRST}
We begin with the full gauge fixed Maxwell action for spin-one, its variation, its BRST variation and BRST variations of all fields 
\besubeqs
\begin{align}
    S&= \int \tfrac14 F_{\mu\nu}^2 +B(\tfrac{\xi}2 B - \pl \cdot \Phi) -\pl_\mu \bar{c} \pl^\mu c\,,\\
    -\delta S&= \int_{\pl M} E_i \delta \Phi^i -B \delta \Phi_z - \delta \bar{c} \pl_z c- \pl_z \bar{c}\delta c\,,\\
    s(\Phi_\mu, c, \bar{c}, B)&= (\pl_\mu c, 0, B,0)\,,\\
    sS&= \int_{\pl M} B \pl_z c\,.
\end{align}
\esubeqs
Suppose we would like to impose Dirichlet boundary conditions, i.e. $\Phi_i$ is fixed, $\delta \Phi_i=0$. The BRST variation gives $s\Phi_i=k_i c$. Therefore, $k_i c|_{\pl M}=0$, i.e. $c|_{\pl M}=0$ and $\delta c|_{\pl M}=0$ (as $k_i$ has no kernel). We should not impose more conditions on $c$ and, hence, we need to impose $B|_{\pl M}=0$ to cancel the BRST variation of the action. From $s\bar{c}=B$ it follows that $\bar{c}|_{\pl M}=0$, which also cancels the rest of the variation. Note that $\xi B=(\pl \cdot \Phi)$, so $(\pl \cdot \Phi)|_{\pl M}=0$ as it can be checked with the Dirichlet propagator. We also see from the gauge condition that $\pl_z \delta\Phi_z|_{\pl M}=0$, hence, $\Phi_z$ obeys Neumann boundary conditions. Indeed, the method of images gives the propagator that satisfies Dirichlet/Neumann boundary conditions for $\Phi_i$/$\Phi_z$. 

The story of Neumann boundary conditions is very similar. The method of images leads to Dirichlet/Neumann boundary conditions for $\Phi_z$/$\Phi_i$. Indeed, we want to fix the electric field $E_i=\pl_z \Phi_i - \pl_i \Phi_z$, i.e. we should fix $\Phi_z$ and $\pl_z\Phi_i$. It follows then $s \Phi_z= \pl_z c$, i.e. $\pl_z c|_{\pl M}=0$. Now, $B$ can be free at the boundary as $\delta \Phi_z=0$. We now need $\pl_z \bar{c}|_{\pl M}=0$ for the variation to vanish, which implies $\pl_z B|_{\pl M}=0$. Again it can be checked that $\pl_z (\pl \cdot \Phi)|_{\pl M}=0$ for the propagator.

\section{Unphysical degrees of freedom}
\label{app:Unphysical}
The physical gauge is very easy to implement conceptually, but more covariant gauges are favored to simplify computations of (especially, higher order) correlators. We need to understand what to do with unphysical degrees of freedom. Let us first consider SDYM as the simplest example. 

In (SD)YM it was very useful to decompose $\Phi^{A,A'}$ into $\Phi_z\sim \epsilon_{CC'}\Phi^{C,C'}$, $\Phi_k\sim k\cdot \Phi$ and the physical components $\Phi_\pm$. Similarly, on the other side we have the Nakanishi-Lautrup field $B\sim \Psi_z\sim \epsilon_{AB}\Psi^{A,B}$, $\Psi_k\sim k\cdot \Psi$ and the physical components $\Psi_\pm$. The free action reads
\begin{align}
    S_2&= \int \Psi_+D_+\Phi_++\Psi_-D_-\Phi_--\Psi_k (\pl_z \Phi_k +k \Phi_z)-\Psi_z(k \Phi_k +\pl_z \Phi_z)\,.
\end{align}
The physical $\Phi_\pm$, $\Psi_\pm$ and unphysical fields (the rest) are not coupled to each other. The equations for $z$- and $k$-components of $\Phi$ and $\Psi$ are coupled and effectively one gets $\square \bullet =0$ for $\Phi_k$, $\Phi_z$, $\Psi_k$, $\Psi_z$. We can diagonalize the equations for $\Phi_k$, $\Phi_z$ and $\Psi_k$, $\Psi_z$ to find
\begin{align}
    D_\pm(\Phi_k\pm\Phi_z)&=0\,, & D_\mp(\Psi_k\pm\Psi_z)&=0\,.
\end{align}
We also see that $\pl_z f_k\sim k f_z$ and $\pl_z f_z\sim kf_k$, where $f\in \{\Phi, \Psi\}$. The latter means that the Dirichlet and Neumann problems are closely related. For example, if we want to impose Dirichlet boundary conditions on $\Phi_k$, then we also fix $\pl_z\Phi_z$, i.e. we have the Neumann problem for $\Phi_z$. As a result we have two types of boundary problems for the unphysical components: Neumann for $\Phi_k, \Psi_z$ and Dirichlet for $\Phi_z, \Psi_k$ and the opposite, i.e. Dirichlet for $\Phi_k, \Psi_z$ and Neumann for $\Phi_z, \Psi_k$.  

The Lagrangian for the unphysical fields have the following general form (below, $p=(\Psi_z,\Psi_k)$ and $q=(\Phi_z, \Phi_k)$)
\begin{align}\label{pqaction}
    \mathcal{L}&= p^T \dot{q} +p^T A q && \rightarrow && p^T \dot{q} +p^T Q^{-1}A Q q && \rightarrow && p^T Kq\,.
\end{align}
With the help of a $GL(2)$-transformation with matrix $Q$ one can bring it to the canonical form to find $K=\mathrm{diag}(D_+,D_-)$. The two-by-two matrix of the most general Green's functions is
\begin{align}\label{matrixprops}
    \begin{pmatrix}
        \langle p_+ q_+\rangle & \langle p_+ q_-\rangle\\
        \langle p_- q_+\rangle & \langle p_- q_-\rangle
    \end{pmatrix}=
    \begin{pmatrix}
        -e^{-k|z'-z|} \theta(z'-z) & 0\\
        c e^{-k(z+z')} & -e^{-k|z-z'|} \theta(z-z')
    \end{pmatrix}\,.
\end{align}
The diagonal carries the solutions to the inhomogeneous equations and the off-diagonal element is the homogeneous solution. Note that the top-right entry has to be zero since there are no regular solutions to $D_-f=0$. For the same reason there are no homogeneous solutions on the diagonal. Only the bottom-left corner has a free coefficient. 

Appendix \ref{app:BRST} recalls what Dirichlet and Neumann problems mean for a gauged fixed gauge theory in the second order formulation, i.e. how to impose boundary conditions on fields and ghosts in a consistent way. Little to nothing changes in the first order formulation of the Maxwell theory. The only change in the SDYM limit is that the equations for the physical components are first order and we can only impose Dirichlet boundary conditions on them. If we do not use the (somewhat nonlocal) momentum space to separate $\Phi_\pm$ from $\Phi_k$, then we should impose Dirichlet on the whole $\Phi_i$, i.e. on $\Phi_k$ as well. As a result, we need $B|_{\pl M}=0$ and $\pl_z \Phi_z|_{\pl M}$ is fixed by the Lorenz gauge condition. This gives 
\begin{align}
    \langle \Psi^{A,B}(-k,z) \Phi^{C,C'}(k,z')\rangle^{\text{F}}&=\langle \Psi^{A,B}(-k,z) \Phi^{C,C'}(k,z')\rangle_{\text{inh}}^{\text{F}} \pm e^{-k(z+z')}\frac{k^A\brk^{B}k^C\brk^{C'}}{4k^2}\,,\\
    \langle \Psi^{A,B}(-k,z) \Phi^{C,C'}(k,z')\rangle_{\text{inh}}^{\text{F}}&=-\frac{1}{2k}e^{-k|z-z'|}\epsilon^{AC} [k^{BC'}-k\epsilon^{BC'}\sign(z-z')]\,.
\end{align}
with the upper sign. Besides $B|_{\pl M}=\Phi_k|_{\pl M}=0$, which fixes it completely, the upper-sign propagator satisfies $\pl_z \Psi_k|_{\pl M}=\pl_z\Phi_z|_{\pl M}=0$. In accordance with the general analysis, the last term is pure gauge, $\nabla^{CC'}e^{-k(z+z')}=e^{-k(z+z')} k^C\brk^{C'}$. Flipping the sign of the last, pure gauge, term exchanges Dirichet/Neumann conditions on the fields $B,\Phi_k$ and $\Psi_k,\Phi_z$. It is this last, lower-sign propagator, that is obtained from Chalmers--Siegel theory starting from Neumann boundary conditions in Feynman gauge, deforming them into mixed ones and, finally, taking the self-dual limit \cite{Skvortsov:2026gtq}. This choice leads to a gauge dependence: the results in the axial and Feynman gauges differ. The difference can be attributed to the Neumann boundary condition since additional terms that deform it to mixed boundary conditions do not contribute to the gauge variation.       

What we observed above is that at the SD-point there is a discrete ambiguity in the choice of the propagator. As a result, there are two propagators. The upper-sign propagator, which can be called Dirichlet, seems better motivated and the results in Feynman and physical gauges agree. Indeed, we fix $\Phi_\pm$ on the boundary and the BRST transformation is $s \Phi_i = k_i c$. Since $k_i$ is considered to be invertible, we have to impose $c|_{\pl M}=0$ and, hence, $B|_{\pl M}=0$ if we treat all components $\Phi_i$ the same way.  

Let us extend these ideas to arbitrary spin. We recall the decomposition of $\Phi^{A(2s-1),A'}$ and $\Psi^{A(2s-1),B}$ (we will ignore the $N$-factor that was found in \eqref{nfactor})
\besubeqs
\begin{align}
    \Phi&=(\bry \pl)\phi+ (\bry y)\rho\,,  & \Psi&=(\bry \pl)\psi+ (\bry y)B\,,\\
    \phi&= q+ (yky) \chi\,, & \psi&=\pi+ (yky)\sigma\,.
\end{align}
\esubeqs
Here $\pi$ and $q$ are transverse, i.e. $(\pl k \pl) \bullet=0$. 
The dictionary with the spin-one case is as follows: $q\sim \Phi_{\pm}$, $\chi\sim \Phi_k$, $\rho\sim \Phi_z$, $\pi\sim \Psi_{\pm}$, $\sigma \sim \Psi_k$ and the Nakanishi-Lautrup field $B$ is the same. The linearized field strength and gauge fixing conditions are
\besubeqs
\begin{align}
    (yFy)=(y\nabla \plb) \Phi&=  \slashed k q +N \dot{q} + (yky)[ \slashed k \chi + (N+2) \dot{\chi} +\rho]\,,\\
    G\equiv (\pl \nabla \plb)\Phi&= -4k^2 (N+1) \chi +(yky)(\pl k\pl) \chi+\slashed k \rho-(N+2)\dot\rho\,.
\end{align}
\esubeqs
Here $\slashed k = (y k \pl)$ and $N=N_y=y^C\pl_C$. The free gauge-fixed action is 
\begin{align}
    S_2&= \langle \pi| \slashed k q +N \dot{q} \rangle + 
    \langle \sigma |(\pl k\pl)(yky)[\slashed k \chi + (N+2) \dot{\chi} +\rho] \rangle + \langle B | G\rangle \,.
\end{align}
The physical components $q$ and $\pi$ decouple. The equations for the unphysical components are more complicated than for $s=1$ ($\slashed k$ vanishes identically on fields without indices). Let us note that the operator $(\pl k\pl)(yky)$ is invertible and we can redefine $\sigma$ to eliminate it. In addition we flip the sign of $B$. Lastly, we rewrite the $\chi$-part of $G$ in terms of $y_k=(ky)$, $y_\brk=(\brk y)$:
\begin{align}
    -4k^2 (N+1) \chi +(yky)(\pl k\pl) \chi&= -4k^2 (N_{y_k}+1)(N_{y_\brk}+1) \chi\,.
\end{align}
The action for the unphysical components is now
\begin{align}\label{unphys}
    S_{\text{unphys}}&= 
    \langle \sigma |\slashed k \chi + (N+2) \dot{\chi} +\rho \rangle + \langle B | 4k^2 (N_{y_k}+1)(N_{y_\brk}+1) \chi+(N+2)\dot\rho-\slashed k \rho \rangle \,.
\end{align}
The action has the familiar form \eqref{pqaction}. To diagonalize the ``potential'' we use $\slashed k= k(N_{y_\brk}-N_{y_k})$. We find $K=(N+2)\mathrm{diag}(D_+,D_-)$, i.e. $(N+2)$ is an overall factor and can be rescaled away with the help of $\sigma$, $B$. Therefore, we find the same two-by-two matrix \eqref{matrixprops} of propagators between $\sigma,B$ and $\chi,\rho$, but with overall identity operator $\epsilon^{AB(2s-2)}$ for indices implied there. Likewise, there is just one parameter, $c$ in \eqref{matrixprops}, to impose boundary conditions.

Note that due to \eqref{unphys} not being diagonal, a Dirichlet condition on, say, $\chi$, implies for $s>1$ a mixed boundary condition for $\rho$ and other way around. Therefore, for $s>1$ we cannot achieve Dirichlet/Neumann conditions for all unphysical fields at the same time, but this is not needed. Finally, we can impose Dirichlet conditions for $\chi$ and $B$, which is what is done in the main text. Flipping the sign of the pure gauge term, one gets Neumann boundary conditions for $\chi$ and $B$. The latter would be inherited if we approach the SD boundary conditions starting from Neumann/mixed conditions in the second order theory in Feynman gauge.

\footnotesize
\providecommand{\href}[2]{#2}\begingroup\raggedright\endgroup

\end{document}